\documentclass[reprint, amsmath, amssymb, superscriptaddress, prb]{revtex4-2}
\usepackage{graphicx}
\usepackage{hyperref}
\usepackage{xcolor}
\usepackage{bbm}

\begin{document}
\title{The scattering coefficients of superconducting microwave resonators:\\
II. System-bath approach}

\author{Qi-Ming Chen}
\email{qiming.chen@wmi.badw.de}
\affiliation{Walther-Mei{\ss}ner-Institut, Bayerische Akademie der Wissenschaften, 85748 Garching, Germany}
\affiliation{Physik-Department, Technische Universit{\"a}t M{\"u}nchen, 85748 Garching, Germany}

\author{Matti Partanen}
\affiliation{Walther-Mei{\ss}ner-Institut, Bayerische Akademie der Wissenschaften, 85748 Garching, Germany}

\author{Florian Fesquet}
\author{Kedar E. Honasoge}
\author{Fabian Kronowetter}
\author{Yuki Nojiri}
\author{Michael Renger}
\author{Kirill G. Fedorov}
\affiliation{Walther-Mei{\ss}ner-Institut, Bayerische Akademie der Wissenschaften, 85748 Garching, Germany}
\affiliation{Physik-Department, Technische Universit{\"a}t M{\"u}nchen, 85748 Garching, Germany}

\author{Achim Marx}
\affiliation{Walther-Mei{\ss}ner-Institut, Bayerische Akademie der Wissenschaften, 85748 Garching, Germany}

\author{Frank Deppe}
\email{frank.deppe@wmi.badw.de}
\author{Rudolf Gross}
\email{rudolf.gross@wmi.badw.de}
\affiliation{Walther-Mei{\ss}ner-Institut, Bayerische Akademie der Wissenschaften, 85748 Garching, Germany}
\affiliation{Physik-Department, Technische Universit{\"a}t M{\"u}nchen, 85748 Garching, Germany}
\affiliation{Munich Center for Quantum Science and Technology (MCQST), Schellingstr. 4, 80799 Munich, Germany}

\date{\today}

\begin{abstract}
	We describe a unified quantum approach for analyzing the scattering coefficients of superconducting microwave resonators with a variety of geometries. We also generalize the method to a chain of resonators with time delays, and reveal several transport properties similar to a photonic crystal. It is shown that both the quantum and classical analyses provide consistent results, and they together reveal different decay and decoherence mechanisms in a general microwave resonator. These results form a solid basis for understanding the scattering spectrums of networks of microwave resonators, and pave the way for applying superconducting microwave resonators in complex circuits.
\end{abstract}

\maketitle

\section{Introduction}
Understanding the scattering coefficients of superconducting microwave resonators is crucial to the study of superconducting quantum circuits \cite{Gu2017}. Owing to the high design flexibility and the strong interactions, a variety of novel photon transport properties emerge when coupling a microwave resonator to other circuit components \cite{Blais2004, Wallraff2004, Shen2005, *Shen2005a, Bufmmodeheckzlseziek1999, Zhou2008, Shen2009, *Shen2009a, Astafiev2010, Liao2010a, Zueco2012, Lalumiere2013, Loo2013, Pichler2016, Huang2013, Li2014, Gu2016, Long2018, Nie2020, Nie2020a}. For example, it is shown that a dissipative atom can completely reflect the photons propagating along a 1D waveguide with no loss \cite{Blais2004, Wallraff2004, Shen2005, *Shen2005a}, although the physical size of the atom is much smaller than the wavelength of the propagating microwave field. Moreover, microwave resonators can also be coupled to each other with different geometries, which lead to many interesting phenomena such as the Fano resonance \cite{Fan2002, Chiba2005, Chak2006, Xiao2010, Longhi2015}, slow light \cite{John1991, Notomi2001, *Notomi2008, Yanik2004, *Yanik2004a, *Yanik2005, Xu2007, Dumeige2009}, coupled-resonator-induced transparency \cite{Smith2004, Totsuka2007, Xiao2007, Yang2009}, and localized bound states \cite{Bulgakov2008, Shi2009, Longo2010, Biondi2014, Biella2015, Qiao2019a, Sundaresan2019}. During the past decades, the scattering coefficients of superconducting quantum circuits have attracted an enormous interest and led to a variety of discoveries. However, most of the existing work either assume the input microwave field to be a few-photon Fock states \cite{Fan2010, *Xu2015} or consider a purely classical microwave field as the input \cite{Yariv1999, *Xu2000}. Moreover, those proposed methods are often limited to certain scenarios without general applicability. For example, one often uses the transfer-matrix method to study a hanger-type resonator, but has to switch to the system-bath description to calculate the scattering coefficients of a necklace-type resonator \cite{Pierre2019, Leppaekangas2019}. A unified approach that applies to a general superconducting microwave resonator is still missing, but it is in high-demand for design larger circuits. 

Here, and also in a parallel paper \cite{Chen2021a}, we study the scattering coefficients of superconducting microwave resonators in either quantum or classical perspectives. In this work, we employ the the system-bath method in quantum optics and derive the analytical descriptions of the scattering coefficients \cite{Collett1984, Gardiner1985, Carmichael2013}. We compare the results with the classical approach \cite{Chen2021a}, and reveal the correspondence of different concepts in the two languages, such as the the damping rates and the quality factors. We also generalize the method to a chain of microwave resonators, which form a photonic-crystal-like system and exhibit interesting transport features that are qualitatively different from a single resonator. These results provide a systematic study of the scattering coefficients of superconducting microwave resonators in the quantum perspective. 

The rest of this paper is organized as follows: In Sec.\,\ref{sec:one_resonator}, we outline the system-bath method and derive the scattering coefficients of a single microwave resonator, which couples to the external circuity in various geometries. Next, we generalize our method to a chain of hanger-type resonators with time delay in Sec.\,\ref{sec:cascade_hanger}, which are side-coupled to an extended waveguide with different time delays. We also study a chain of necklace-type resonators in Sec.\,\ref{sec:cascade_necklace}, which are coupled to each other through the ends. Finally, we conclude this study, compare the classical and quantum methods, and discuss how the dephasing effect can be incorporated in the analysis in Sec.\,\ref{sec:conclusions}. Detailed derivations of the input-output relations and the scattering coefficients for resonators with different geometries can be found in Appendices\,\ref{app:input_output}-\ref{app:cascade_necklace}.

\section{scattering coefficients of a single microwave resonator} \label{sec:one_resonator}
\subsection{The hanger-type $\lambda/4$ resonators} \label{sec:hanger}
\begin{figure}
\centering
\includegraphics[width=0.9\columnwidth]{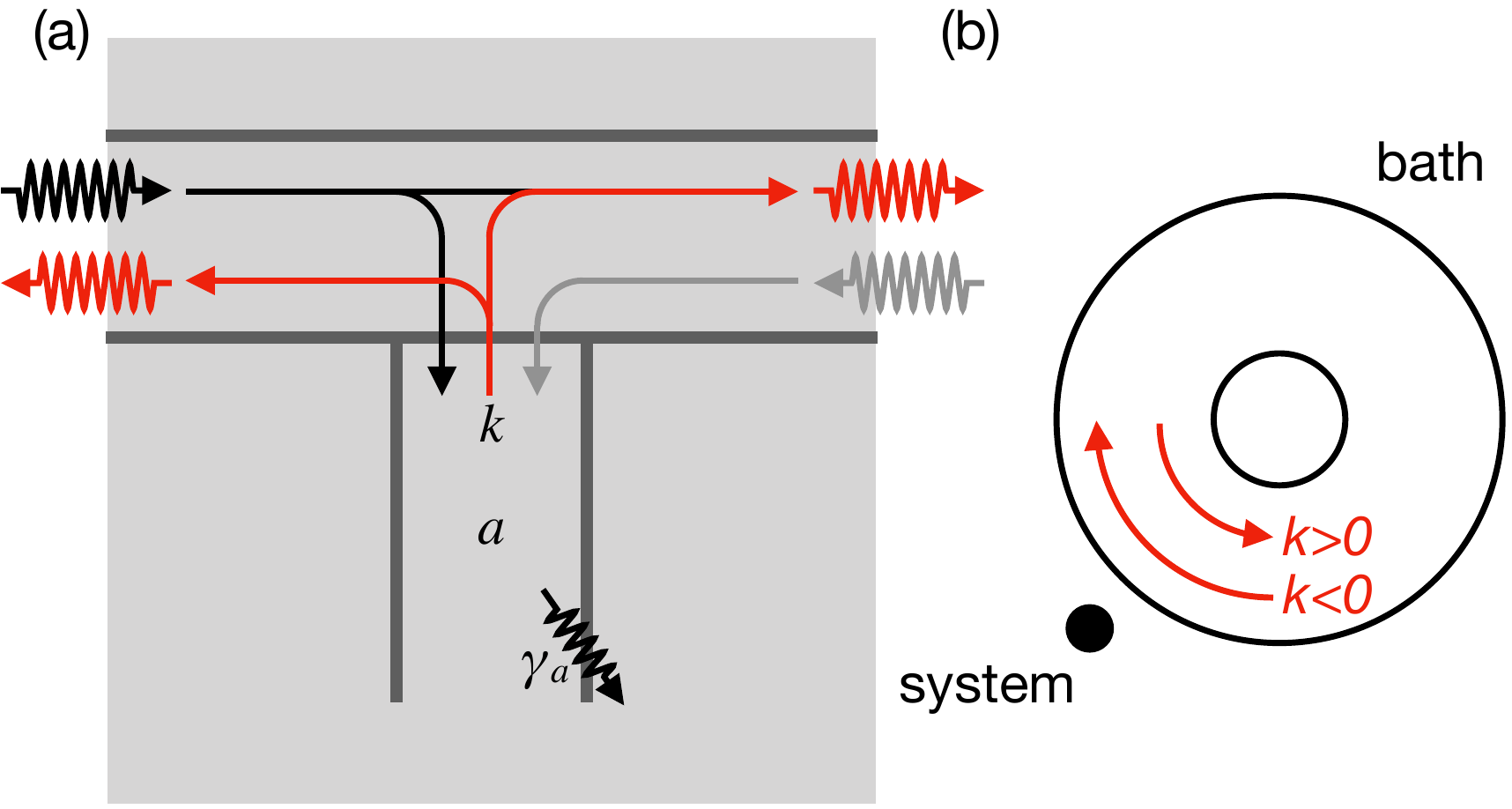}
\caption{Schematic of a hanger-type $\lambda/4$ resonator. Here, a short-circuited $\lambda/4$ (i.e., the system) is side-coupled to a transmission-line waveguide (i.e., the bath), which accommodates both left- and right-propagating fields. Panel (b) shows a simplified schematic of (a), where we describe the bath as a circular tube indicating the use of periodic boundary conditions.}
\label{fig:hanger}
\end{figure}

The hanger-type $\lambda/4$ resonator is schematically shown in Fig.\,\ref{fig:hanger}(a)-(b), where the intra-resonator field, $a$, is coupled to the modes $b$ of an waveguide with a coupling strength $\kappa$. We describe the coupled system as 
\begin{align}
	H_{\rm s} &= \hbar \omega_{0} a^{\dagger}a, \\
	H_{\rm b} &=\sum_{k=-\infty}^{+\infty} \hbar \omega_k b_{k}^{\dagger} b_{k}, \\
	H_{\rm sb} &= \sum_{k=-\infty}^{+\infty} \hbar
	\left(\kappa^{*} a b_{k}^{\dagger}  
	+ \kappa a^{\dagger} b_{k} \right).
\end{align}
Following the standard procedure of the input-output analysis \cite{Collett1984, Gardiner1985}, which is also outlined in Appendix\,\ref{app:input_output}, one can derive a set of linear equations that relate the dynamics of the intra-resonator field, $a$, to a bath of input and output fields, $b$. However, the input and output fields defined in this way are not directly related to the fields measured in experiments. To calculate the transport properties of the system, one must distinguish the left and the right propagating fields, which we denote as $l$ and $r$, from the bath $b$, as shown in Fig.\,\ref{fig:hanger}(b). 

Keeping this in mind, we constrain our discussion to a small frequency interval around the central driving frequency, $\omega_{\rm d}$. Within this small interval we can approximate the dispersion by the linear relation \cite{Chak2006}
\begin{align}
	\omega_k = \omega_{\rm d} \mp v_{\rm g}\delta_k.
\end{align}
Here, $v_{\rm g}$ is the group velocity of the waveguide, $\delta_k = k \pm k_{\rm d}$ with $k$ and $k_{\rm d}>0$ being the wave vectors that correspond to the frequencies $\omega_k$ and $\omega_{\rm d}$, respectively. In the rotating frame at $\omega_{\rm d}$, we obtain an equivalent description of the coupled system 
\begin{align}
	H_{\rm s} &= \hbar \left(\omega_{0}-\omega_{\rm d} \right) a^{\dagger}a, \\
	H_{\rm b} &= \sum_{\Delta_k=-\infty}^{+\infty} \hbar v_{\rm g}\Delta_{k}\left(-l_{k_{\rm r}-\Delta_{k}}^{\dagger} l_{k_{\rm r}-\Delta_{k}} 
	+ r_{k_{\rm r}+\Delta_{k}}^{\dagger} r_{k_{\rm r}+\Delta_{k}}\right), \\
	H_{\rm sb} &= \sum_{\Delta_k=-\infty}^{+\infty} \hbar
	\left(\kappa^{*} a l_{k_{\rm r}-\Delta_{k}}^{\dagger}  
	+ \kappa a^{\dagger} l_{k_{\rm r}-\Delta_{k}} \right. \nonumber \\
	&\,\,\,\,\,\,\,\,\,\,\,\,
	+ \left. \kappa^{*} a r_{k_{\rm r}+\Delta_{k}}^{\dagger}  
	+ \kappa a^{\dagger} r_{k_{\rm r}+\Delta_{k}} \right).
\end{align}
Here, we have extended the upper and lower limits of the summation to infinity for mathematical convenience, which is valid as long as $\omega_{\rm d}$ is much larger than the typical bandwidth of interest \cite{Fan2010, *Xu2015}. Finally, we complete our transformation by defining $\omega = v_{\rm g}\delta_{k}$, $\Delta_{a}=\omega_{0}-\omega_{\rm d}$, and replacing the discrete field operators by a continuum: $l_{k_{\rm r}-\Delta_{\rm k}} \rightarrow l_{\omega}$, $r_{k_{\rm r}+\Delta_{\rm k}} \rightarrow r_{\omega}$. The result is
\begin{align}
	H_{\rm s} &= \hbar \Delta_{a} a^{\dagger}a, \\
	H_{\rm b} &= \int_{-\infty}^{+\infty}d\omega \hbar \omega \left(-l_{\omega}^{\dagger} l_{\omega}
	+ r_{\omega}^{\dagger} r_{\omega}\right), \\
	H_{\rm sb} &= \int_{-\infty}^{+\infty}d\omega \hbar
	\left(\kappa^{*} a l_{\omega}^{\dagger}  
	+ \kappa a^{\dagger} l_{\omega} 
	+ \kappa^{*} a r_{\omega}^{\dagger}  
	+ \kappa a^{\dagger} r_{\omega} \right).
\end{align}

By separating the left- and right-propagating fields in the waveguide, we split the original bath, $b$, into two independent baths, $l$ and $r$, representing different directions of field propagation in the 1D waveguide. In this way, the scattering coefficients of the resonator can be readily obtained by following the standard input-output analysis. Here, we list several major steps for illustration. Using the Heisenberg equations of motion, we describe the dynamics of the intra-resonator field, $a$, and the two bath fields, $l_{\omega}$ and $r_{\omega}$, as 
\begin{align}
	\dot{a} &= -i\Delta_{a}a 
	-i\int_{-\infty}^{+\infty}d\omega \kappa\left(l_{\omega}+r_{\omega}\right), \label{eq:hanger4_dynamics_a} \\
	\dot{l}_{\omega} &= +i\omega l_{\omega} - i\kappa^{*}a,  \label{eq:hanger4_dynamics_l} \\
	\dot{r}_{\omega} &= -i\omega r_{\omega} - i\kappa^{*}a.  \label{eq:hanger4_dynamics_r}
\end{align}
We further define the input and output fields corresponding to the two baths as
\begin{align}
	l_{\rm in} &= \frac{1}{\sqrt{2\pi}}\int_{-\infty}^{+\infty} d\omega e^{+i\omega t}l_{\omega},\,
	l_{\rm out} = l_{\rm in} + \sqrt{\gamma}a,  \label{eq:hanger4_in_out_l}\\
	r_{\rm in} &= \frac{1}{\sqrt{2\pi}}\int_{-\infty}^{+\infty} d\omega e^{-i\omega t}r_{\omega},\,
	r_{\rm out} = r_{\rm in} + \sqrt{\gamma}a, \label{eq:hanger4_in_out_r}
\end{align}
where $\sqrt{\gamma} = i\sqrt{2\pi}\kappa$. Inserting Eqs.\,\eqref{eq:hanger4_in_out_l}-\eqref{eq:hanger4_in_out_r} into \eqref{eq:hanger4_dynamics_a}-\eqref{eq:hanger4_dynamics_r}, we obtain 
\begin{align}
	\dot{a} &= -i\Delta_{a}a 
	- \left(\gamma+\frac{\gamma_{a}}{2}\right) a 
	- \sqrt{\gamma}\left(l_{\rm in}+r_{\rm in}\right). \label{eq:hanger4_in_out_a}
\end{align}
Here, we have added the intrinsic damping of the resonator, $\gamma_{a}/2$, by hand. 

Equations\,\eqref{eq:hanger4_in_out_l}-\eqref{eq:hanger4_in_out_r} and \eqref{eq:hanger4_in_out_a} determine the scattering coefficients of a hanger-type $\lambda/4$ resonator, which are defined as
\begin{align}
	S_{11} &= \frac{\langle l_{\rm out} \rangle}{\langle r_{\rm in}\rangle},\,
	S_{21} = \frac{\langle r_{\rm out} \rangle}{\langle r_{\rm in} \rangle} 
	\,\text{with}\,\langle l_{\rm in}\rangle = 0, \label{eq:scattering_responses_hanger_1} \\
	S_{12} &= \frac{\langle l_{\rm out} \rangle}{\langle l_{\rm in} \rangle},\,
	S_{22} = \frac{\langle r_{\rm out} \rangle}{\langle l_{\rm in}\rangle}
	\,\text{with}\,\langle r_{\rm in}\rangle = 0. \label{eq:scattering_responses_hanger_2}
\end{align}
That is, 
\begin{align}
	S_{11} &= S_{22} = - \frac{\gamma}{i\Delta_a + \left(\gamma+\frac{\gamma_a}{2}\right)}, \label{eq:quantum_hanger_1} \\
	S_{21} &= S_{12} = 1 - \frac{\gamma}{i\Delta_a + \left(\gamma+\frac{\gamma_a}{2}\right)}. \label{eq:quantum_hanger_2}
\end{align}
We recall that the scattering coefficients of a hanger-type $\lambda/4$ resonator which are derived by using the transfer-matrix approach are given by \cite{Chen2021a}
\begin{align}
	S_{11} = S_{22} &\approx -\frac{\frac{\omega_{a}}{2Q_{\rm c}}}{i\Delta_{a} 
	+ \left(\frac{\omega_{a}}{2Q_{\rm i}}+\frac{\omega_{a}}{2Q_{\rm c}}\right)}, \label{eq:circuit_hanger_1}\\
	S_{21} = S_{12} &\approx 1 -\frac{\frac{\omega_{a}}{2Q_{\rm c}}}{i\Delta_{a} 
	+ \left(\frac{\omega_{a}}{2Q_{\rm i}}+\frac{\omega_{a}}{2Q_{\rm c}}\right)}. \label{eq:circuit_hanger_2}
\end{align}
Here, we have replaced the imaginary unit, $j$, that follows the convension of microwave engineering by the imaginary unit $i=-j$ \cite{Girvin2011}. Comparing Eqs.\,\eqref{eq:quantum_hanger_1}-\eqref{eq:quantum_hanger_2} with \eqref{eq:circuit_hanger_1}-\eqref{eq:circuit_hanger_2}, we obtain the following relations between the damping rates and the quality factors:
\begin{align}
	\gamma_{a} = \frac{\omega_{a}}{Q_{\rm i}},\
	\gamma = \frac{\omega_{a}}{2Q_{\rm c}}.
\end{align}
The above relation also holds for a hanger-type $\lambda/2$ resonator, but with different definitions of the resonant frequency and the quality factors. The physical interpretation of this result is obvious: First, if the coupling rate between the resonator and the waveguide vanishes ($\gamma \rightarrow 0$), the external quality factor diverges ($Q_c \rightarrow \infty$). Second, if the intrinsic decay rate of the resonator vanishes ($\gamma_a \rightarrow 0$), the internal quality factors diverges ($Q_i \rightarrow \infty$). We note that the factor of \textit{two} in the expression of $\gamma$ originates from the fact that both the left- and right-propagating fields in the waveguide couple to the intra-resonator field. However, the effective energy decay rate in both the internal and coupling dissipation channels equals to the ratio between the resonant frequency and the corresponding Q factor. 

\subsection{The necklace-type $\lambda/2$  resonator} \label{sec:transmission}
\begin{figure}
\centering
\includegraphics[width=0.9\columnwidth]{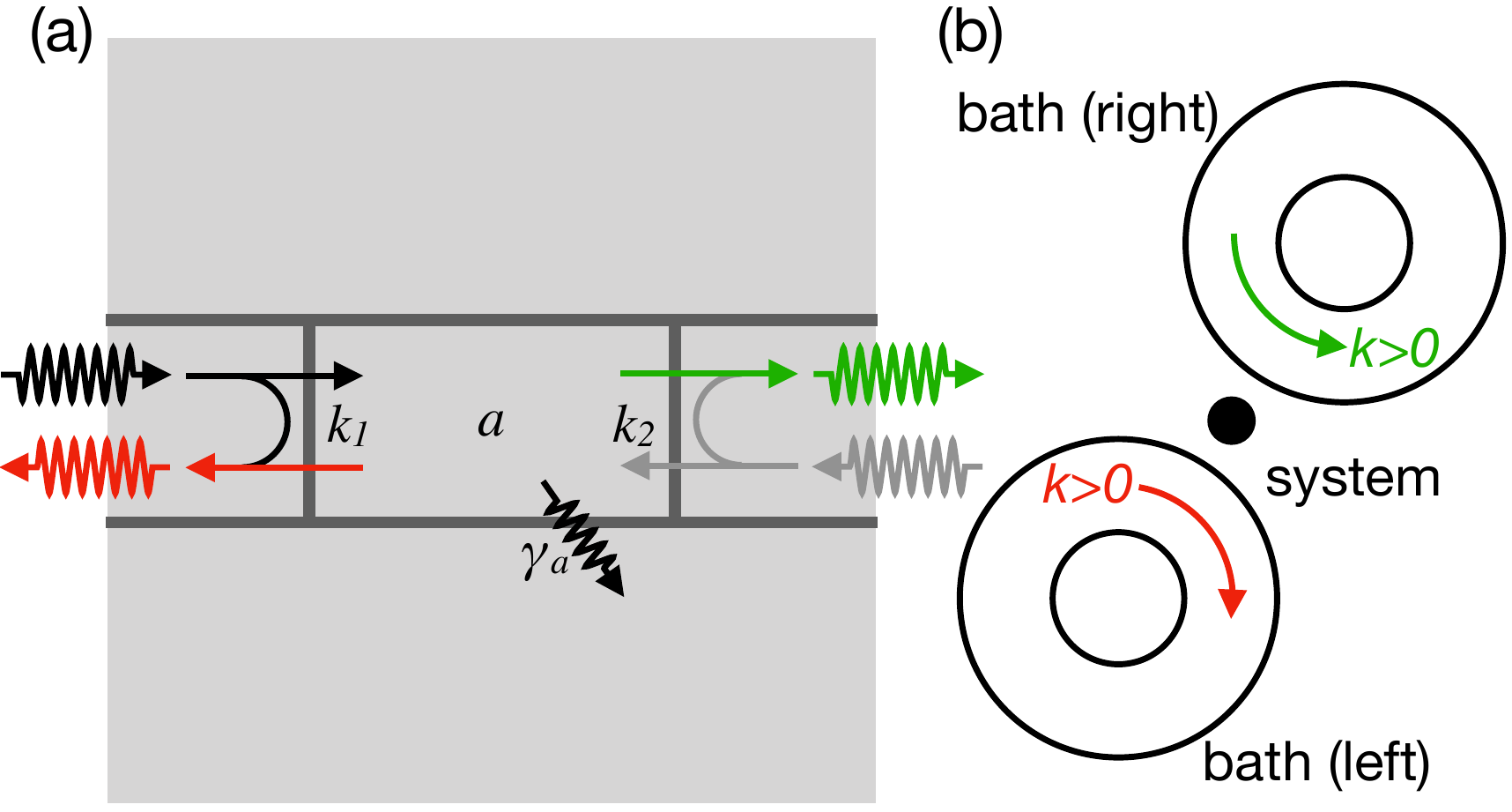}
\caption{Schematic of a necklace-type $\lambda/2$ resonator. Here, an open-circuited $\lambda/2$ (i.e., the system) is directly coupled to two transmission-line waveguides (i.e., the baths) at the two ends, respectively. Each of the bath accommodates only one unidirectional propagating field. We use different colors for the outgoing fields emitting to emphasize the $\pi$ phase shift of the intra-resonator spatial mode at the two ends.}
\label{fig:necklace}
\end{figure}

The necklace-type $\lambda/2$ resonator is schematically shown in Fig.\,\ref{fig:necklace}(a)-(b), where the intra-resonator field, $a$, is coupled to two independent baths, $b_{1}$ and $b_{2}$, on the left and right hand side, respectively. We describe the whole system as
\begin{align}
	H_{\rm s} &= \hbar \omega_{\rm d} a^{\dagger}a, \\
	H_{\rm b} &= \sum_{m=1}^{2}\sum_{k=0}^{+\infty} \hbar \omega_k b_{m,k}^{\dagger} b_{m,k}, \\
	H_{\rm sb} &= \sum_{m=1}^{2}\left(-1\right)^{m-1}\sum_{k=0}^{+\infty} \hbar
	\left(\kappa_{m}^{*} a b_{m,k}^{\dagger}  
	+ \kappa_{m}a^{\dagger} b_{m,k} \right). \label{eq:necklace2_sr}
\end{align}
Here, the wave vector, $k$, takes only positive values that defines a unidirectional propagation of the microwave fields in the two feedlines, as shown in Fig.\,\ref{fig:necklace}(b). The phase factor $\pm 1$ in the system-bath interaction takes into account the $\pi$ phase difference of the spatial modes with $(2n+1) \lambda/2 =l$ at the two ends of the resonator, whereas there is no phase difference for the modes with $2n \lambda/2 =l$. 

Following a similar procedure as before, we linearize the dispersion relation around the central driving frequency
\begin{align}
	\omega_{k}=\omega_{\rm d}+v_{\rm g}\Delta_{k}.
\end{align} 
Eventually, we obtain the Hamiltonian in terms of photon frequencies
\begin{align}
	H_{\rm s} &= \hbar \Delta_{a} a^{\dagger}a, \\
	H_{\rm b} &= \sum_{m=1}^{2}\int_{-\infty}^{+\infty}d\omega \hbar \omega b_{m,\omega}^{\dagger} b_{m,\omega}, \\
	H_{\rm sb} &= \sum_{m=1}^{2}\left(-1\right)^{m-1}\int_{-\infty}^{+\infty}d\omega \hbar
	\left(\kappa_{m}^{*} a b_{m,\omega}^{\dagger}  
	+ \kappa_{m} a^{\dagger} b_{m,\omega}  \right).
\end{align}
By using the Heisenberg equations of motion, we describe the dynamics of the intra-resonator field, $a$, and the two bath fields, $b_{1,\omega}$ and $b_{2,\omega}$, as
\begin{align}
	\dot{a} &= -i\Delta_{a}a 
	-i\sum_{m=1}^{2}\left(-1\right)^{m-1}\int_{-\infty}^{+\infty}d\omega \kappa_{m} b_{m,\omega}, \label{eq:necklace2_dynamics_a} \\
	\dot{b}_{1,\omega} &= -i\omega b_{1,\omega} - i\kappa_{1}^{*}a,  \label{eq:necklace2_dynamics_b1} \\
	\dot{b}_{2,\omega} &= -i\omega b_{2,\omega} + i\kappa_{2}^{*}a.  \label{eq:necklace2_dynamics_b2}
\end{align}
We further define the input and output fields corresponding to the two baths as
\begin{align}
	b_{1,{\rm in}} &= \frac{1}{\sqrt{2\pi}}\int_{-\infty}^{+\infty} d\omega e^{+i\omega t}b_{1,\omega},\,
	b_{1,{\rm out}} = b_{1,{\rm in}} + \sqrt{\gamma_{1}}a,  \label{eq:necklace2_in_out_b1}\\
	b_{2,{\rm in}} &= \frac{1}{\sqrt{2\pi}}\int_{-\infty}^{+\infty} d\omega e^{-i\omega t}b_{2,{\omega}},\,
	b_{2,{\rm out}} = b_{2,{\rm in}} - \sqrt{\gamma_{2}}a. \label{eq:necklace2_in_out_b2}
\end{align}
Inserting Eqs.\,\eqref{eq:necklace2_in_out_b1}-\eqref{eq:necklace2_in_out_b2} into \eqref{eq:necklace2_dynamics_a}-\eqref{eq:necklace2_dynamics_b2}, we obtain 
\begin{align}
	\dot{a} &= -i\Delta_{a}a 
	- \left(\frac{\gamma_{1}+\gamma_{2}}{2}+\frac{\gamma_{a}}{2}\right) a 
	+ \sum_{m=1}^{2}\left(-1\right)^{m}\sqrt{\gamma_{m}}b_{m,{\rm in}}. \label{eq:necklace2_in_out_a}
\end{align}
Here, we have defined $\sqrt{\gamma_{m}} = i\sqrt{2\pi}\kappa_{m}$ and added the intrinsic damping of the resonator, $\gamma_{a}/2$, by hand. 

Equations\,\eqref{eq:necklace2_in_out_b1}-\eqref{eq:necklace2_in_out_b2} and \eqref{eq:necklace2_in_out_a} determine the scattering coefficients of a necklace-type $\lambda/2$ resonator, which are defined as
\begin{align}
	S_{11} &= \frac{\langle b_{1,{\rm out}} \rangle}{\langle b_{1,{\rm in}}\rangle},\,
	S_{21} = \frac{\langle b_{2,{\rm out}} \rangle}{\langle b_{1,{\rm in}}\rangle} 
	\,\text{with}\,\langle b_{2,{\rm in}}\rangle = 0, \label{eq:scattering_responses_necklace_1} \\
	S_{12} &= \frac{\langle b_{1,{\rm out}} \rangle}{\langle b_{2,{\rm in}}\rangle},\,
	S_{22} = \frac{\langle b_{2,{\rm out}} \rangle}{\langle b_{2,{\rm in}}\rangle}
	\,\text{with}\,,\langle b_{1,{\rm in}}\rangle = 0. \label{eq:scattering_responses_necklace_2}
\end{align}
That is,
\begin{align}
	S_{11} &= 1 - \frac{\gamma_{\rm l}}{i\Delta_a + \left(\frac{\gamma_{\rm l}+\gamma_{\rm b}}{2}+\frac{\gamma_a}{2}\right)}, \label{eq:quantum_transmission_1} \\
	S_{21} &= S_{12} = \frac{\sqrt{\gamma_{\rm l}\gamma_{\rm b}}}{i\Delta_a + \left(\frac{\gamma_{\rm l}+\gamma_{\rm b}}{2}+\frac{\gamma_a}{2}\right)}, \\
	S_{22} &= 1 - \frac{\gamma_{\rm b}}{i\Delta_a + \left(\frac{\gamma_{\rm l}+\gamma_{\rm b}}{2}+\frac{\gamma_a}{2}\right)}. \label{eq:quantum_transmission_4}
\end{align}
We recall that the scattering coefficients of a necklace-type $\lambda/2$ resonator which have been derived by using the transfer-matrix approach \cite{Chen2021a}
\begin{align}
	S_{11} &\approx 1- \frac{\frac{\omega_{a}}{Q_{\rm c_1}}}{i\Delta_{a} 
	+ \left(\frac{\omega_{a}}{2Q_{\rm i}}+\frac{\omega_{a}}{2Q_{\rm c_1}}+\frac{\omega_{a}}{2Q_{\rm c_2}}\right)}, \label{eq:circuit_transmission_1}\\
	S_{21} &= S_{12} \approx \frac{\frac{\omega_{a}}{\sqrt{Q_{\rm c_1}Q_{\rm c_2}}}}{i\Delta_{a} 
	+ \left(\frac{\omega_{a}}{2Q_{\rm i}}+\frac{\omega_{a}}{2Q_{\rm c_1}}+\frac{\omega_{a}}{2Q_{\rm c_2}}\right)}, \label{eq:circuit_transmission_2}\\
	S_{22} &\approx 1- \frac{\frac{\omega_{a}}{Q_{\rm c_2}}}{i\Delta_{a} 
	+ \left(\frac{\omega_{a}}{2Q_{\rm i}}+\frac{\omega_{a}}{2Q_{\rm c_1}}+\frac{\omega_{a}}{2Q_{\rm c_2}}\right)}.  \label{eq:circuit_transmission_3}
\end{align}
Comparing Eqs.\eqref{eq:quantum_transmission_1}-\eqref{eq:quantum_transmission_4}, with \eqref{eq:circuit_transmission_1}-\eqref{eq:circuit_transmission_3},  we obtain the following relations between the decay rates and the quality factos of a necklace-type $\lambda/2$ resonator
\begin{align}
	\gamma_a &= \frac{\omega_{a}}{Q_{i}},\,
	\gamma_{\rm 1} = \frac{\omega_{a}}{Q_{c_1}},\
	\gamma_{\rm 2} = \frac{\omega_{a}}{Q_{c_2}},
\end{align}
which also hold for a necklace-type $\lambda/4$ resonator but with different definitions of the resonant frequency and the quality factors.

\subsection{The bridge-type $\lambda/2$ resonators} \label{sec:hanger}
\begin{figure}
\centering
\includegraphics[width=0.9\columnwidth]{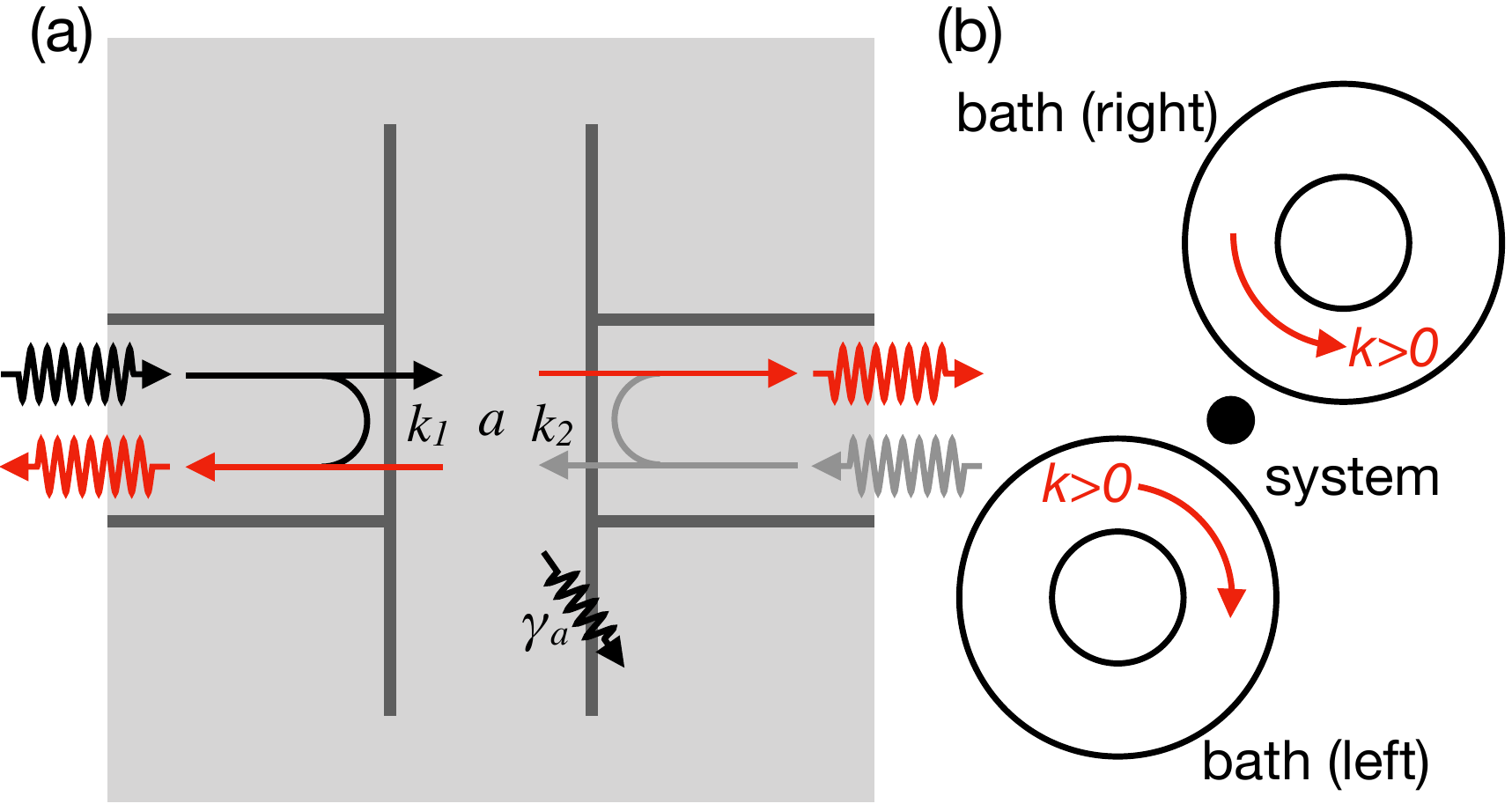}
\caption{Schematic of a bridge-type $\lambda/2$ resonator. Here, a short-circuited $\lambda/2$ (i.e., the system) is directly coupled to two transmission-line waveguides (i.e., the baths) at the same anti-node. Each of the bath accommodates only one unidirectional propagating field.}
\label{fig:cross}
\end{figure}

The bridge-type $\lambda/2$ resonator is schematically shown in Fig.\,\ref{fig:cross}(a)-(b), where the intra-resonator field, $a$, is coupled to two independent baths, $b_{1}$ and $b_{2}$, on the left and right sides, respectively. Compared with the Hamiltonian for a necklace-type $\lambda/2$ resonator, the major difference lies in the system-bath interaction term
\begin{align}
	H_{\rm sb} &= \sum_{m=1}^{2}\sum_{k=0}^{+\infty} \hbar
	\left(\kappa_{m}^{*} a b_{m,k}^{\dagger}  
	+ \kappa_{m}a^{\dagger} b_{m,k} \right).
\end{align}
That is, we assume no phase difference in the system-bath interaction as the two baths are coupled to the resonator mode at the same voltage anti-node. Consequently, the input-output relations are also similar to that of the necklace-type $\lambda/2$ resonator 
\begin{align}
	&b_{1,{\rm out}} = b_{1,{\rm in}} + \sqrt{\gamma_{1}}a,  \label{eq:cross2_in_out_b1} \\
	&b_{2,{\rm out}} = b_{2,{\rm in}} + \sqrt{\gamma_{2}}a. \label{eq:cross2_in_out_b2} \\
	&\dot{a} = -i\Delta_{a}a 
	- \left(\frac{\gamma_{1}+\gamma_{2}}{2}+\frac{\gamma_{a}}{2}\right) a 
	- \sum_{m=1}^{2}\sqrt{\gamma_{m}}b_{m,{\rm in}}, \label{eq:cross2_in_out_a}
\end{align}
Here, we have also added the intrinsic damping of the resonator, $\gamma_{a}/2$, by hand. The scattering coefficients of a bridge-type $\lambda/2$ resonator are 
\begin{align}
	S_{11} &= 1 - \frac{\gamma_{\rm l}}{i\Delta_a + \left(\frac{\gamma_{\rm l}+\gamma_{\rm b}}{2}+\frac{\gamma_a}{2}\right)}, \label{eq:quantum_cross2_s11} \\
	S_{21} &= S_{12} = -\frac{\sqrt{\gamma_{\rm l}\gamma_{\rm b}}}{i\Delta_a + \left(\frac{\gamma_{\rm l}+\gamma_{\rm b}}{2}+\frac{\gamma_a}{2}\right)}, \\
	S_{22} &= 1 - \frac{\gamma_{\rm b}}{i\Delta_a + \left(\frac{\gamma_{\rm l}+\gamma_{\rm b}}{2}+\frac{\gamma_a}{2}\right)}. \label{eq:quantum_cross2_s22}
\end{align}
We recall that the scattering coefficients of a bridge-type $\lambda/2$ resonator which have been derived by using the classical transfer-matrix approach \cite{Chen2021a}
\begin{align}
	S_{11} &\approx 1- \frac{\frac{\omega_{a}}{Q_{\rm c_1}}}{i\Delta_{a} 
	+ \left(\frac{\omega_{a}}{2Q_{\rm i}}+\frac{\omega_{a}}{2Q_{\rm c_1}}+\frac{\omega_{a}}{2Q_{\rm c_2}}\right)}, \label{eq:circuit_cross2_s11}\\
	S_{21} &= S_{12} \approx \frac{\frac{\omega_{a}}{\sqrt{Q_{\rm c_1}Q_{\rm c_2}}}}{i\Delta_{a} 
	+ \left(\frac{\omega_{a}}{2Q_{\rm i}}+\frac{\omega_{a}}{2Q_{\rm c_1}}+\frac{\omega_{a}}{2Q_{\rm c_2}}\right)}, \label{eq:circuit_cross2_s21}\\
	S_{22} &\approx 1- \frac{\frac{\omega_{a}}{Q_{\rm c_2}}}{i\Delta_{a} 
	+ \left(\frac{\omega_{a}}{2Q_{\rm i}}+\frac{\omega_{a}}{2Q_{\rm c_1}}+\frac{\omega_{a}}{2Q_{\rm c_2}}\right)}.  \label{eq:circuit_cross2_s22}
\end{align}Comparing Eqs.\,\eqref{eq:quantum_cross2_s11}-\eqref{eq:quantum_cross2_s22} with \eqref{eq:circuit_cross2_s11}-\eqref{eq:circuit_cross2_s22}, we obtain the following relations between the decay rates and the quality factos of a bridge-type $\lambda/2$ resonator
\begin{align}
	\gamma_a &= \frac{\omega_{a}}{Q_{i}},\,
	\gamma_{\rm 1} = \frac{\omega_{a}}{Q_{c_1}},\,
	\gamma_{\rm 2} = \frac{\omega_{a}}{Q_{c_2}}.
\end{align}

As a final remark we emphasize that the system-bath method provides an elegant and unified approach for deriving the scattering coefficients of a general microwave resonator. Here, the key idea is to linearize the dispersion relation of the waveguide and transform the Hamiltonian from the wave vector space to the frequency space. This separation is natural for necklace- and bridge-type resonators with two spatially separated baths, and has been reported recently in the literature \cite{Pierre2019, Leppaekangas2019}. However, it is not a trivial task to apply the method to a hanger-type resonator. Here, we artificially split the single physical bath into two baths with opposite signs of the wave vectors \cite{Chak2006}, and relate the scattering coefficients to these two new fields. Depending on the specific geometry of the system, the scattering coefficients can be readily obtained by taking the mean value of the field operators in Eqs.\,\eqref{eq:scattering_responses_hanger_1}-\eqref{eq:scattering_responses_hanger_2} for hanger-type resonators, or in Eqs.\,\eqref{eq:scattering_responses_necklace_1}-\eqref{eq:scattering_responses_necklace_2} for necklace- or bridge-type resonators.

\section{Coupling multiple hanger-type resonators to a long waveguide} \label{sec:cascade_hanger}
\subsection{General scattering coefficients}
\begin{figure}[ht]
\centering
\includegraphics[width=0.9\columnwidth]{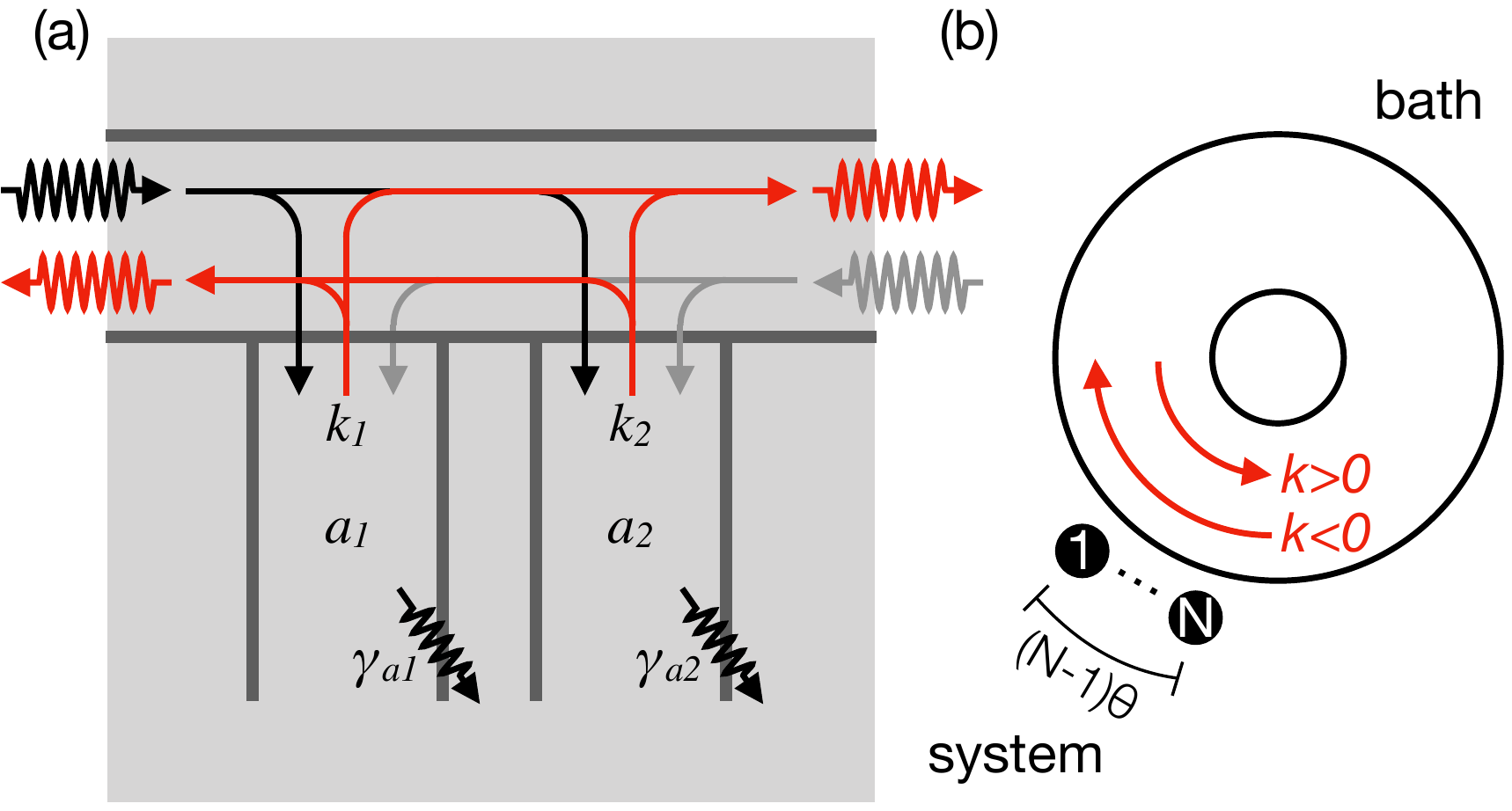}
\caption{Schematic of a chain of $N$ hanger-type $\lambda/4$ resonators (i.e., the system) that are side-coupled to a long transmission-line waveguide (i.e., the bath). The bath accommodates both left- and right-propagating fields. We denote the time delay between a propagating photon and the different short-circuited $\lambda/4$ resonators as $\theta$.}
\label{fig:cascade_hanger}
\end{figure}

Let us now consider a more complex system with $N$ hanger-type resonators that are side-coupled to a long waveguide, as schematically shown in Fig.\,\ref{fig:cascade_hanger}(a)-(b). The total Hamiltonian is
\begin{align}
	H_{\rm s} &= \sum_{j=1}^{N}\hbar \Delta_{j} a_{j}^{\dagger}(t)a_{j}(t), \\
	H_{\rm b} &= \int_{-\infty}^{+\infty}d\omega \hbar \omega \left[-l_{\omega}^{\dagger}(t)l_{\omega}(t) + r_{\omega}^{\dagger}(t)r_{\omega}(t)\right], \\
	H_{\rm sb} &= \sum_{j=1}^{N}\int_{-\infty}^{+\infty}d\omega \hbar
	\left\{ e^{-i(j-1)\omega\tau}\kappa_{j}^{*} a_{j}(t) l_{\omega}^{\dagger}\left[t-(j-1)\tau\right] \right. \nonumber \\ 
	&+ \left. \kappa_{j}^{*} a_{j}(t) r_{\omega}^{\dagger}\left[t-\left(j-1\right)\tau\right]
	+ \text{c.c.} \right\}.
\end{align}
Here, we explicitly included the phase differences $\theta_j = (j-1)\omega\tau = (j-1)\theta$ in the interaction between the bath fields, $l_{\omega}$ and $r_{\omega}$, and the different resonators, $a_{j}$, which are coupled to the waveguide at different positions labelled by $j$. Following the derivations in Appendix\,\ref{app:cascade_hanger}, we obtain the following relation that describes the evolution of the intra-resonator field
\begin{align}
	\dot{a}_{j} &= -i\omega_{j}a_{j} - \frac{\gamma_{a_j}}{2}a_{j}
	-\sum_{j'=1}^{N}\sqrt{\gamma_{j}}\left(\sqrt{\gamma_{j'}}\right)^{*}e^{\left|j'-j\right|\theta} a_{j'}(t) \nonumber \\
	&- \sqrt{\gamma_j} l_{\rm in}\left(t\right)e^{-i(j-N)\theta}
	- \sqrt{\gamma_j} r_{\rm in}\left(t\right)e^{i(j-1)\theta}. 
\end{align}
Here, $\theta=\omega_{\rm d}\tau$, with $\omega_{\rm d}$ being the central driving frequency, is the phase difference in the coupling rate between two neighboring resonators, and $\tau$ the time delay of the propagating field traveling between the finite distance between them. The operators, $l_{\rm in}$ and $r_{\rm in}$, are respectively defined as the input fields at the right and left hand side of the waveguide, which propagate in opposite directions, as shown in Fig.\,\ref{fig:cascade_hanger}(b). Correspondingly, we define $l_{\rm out}$ and $r_{\rm out}$ as the output fields at the left and right hand side of the waveguide, respectively
\begin{align}
	l_{\rm out} =  e^{i(N-1)\theta}l_{\rm in} +\sum_{j'=1}^{N} \left(\sqrt{\gamma_{j'}}\right)^{*}e^{i(j'-1)\theta}a_{j}, \\
	r_{\rm out} =  e^{i(N-1)\theta}r_{\rm in} + \sum_{j'=1}^{N} \left(\sqrt{\gamma_{j'}}\right)^{*}e^{i(N-j')\theta}a_{j}.
\end{align}

With these results, the scattering coefficients of the whole system can be readily obtained by using the expression in Eqs.\,\eqref{eq:scattering_responses_hanger_1}-\eqref{eq:scattering_responses_hanger_2}. For example, for the simplest case with $N=2$ we have
\begin{widetext}
\begin{align}
	S_{11} &= -\frac{\gamma_1\left(i\Delta_{2} + \gamma_{a_2}\right)
	+e^{i2\theta}\gamma_2\left(i\Delta_{1} + \gamma_{a_1}\right)
	+\gamma_1\gamma_2\left(1-e^{i2\theta}\right)}
	{\left(i\Delta_{1} + \gamma_{a_1}\right)
	\left(i\Delta_{2} + \gamma_{a_2}\right)
	+ \gamma_1\left(i\Delta_{2} + \gamma_{a_2}\right)
	+ \gamma_2\left(i\Delta_{1} + \gamma_{a_1}\right)
	+ \gamma_1\gamma_2\left(1-e^{i2\theta}\right)}, \\
	S_{21} &= S_{12} = \frac{e^{i\theta}\left(i\Delta_{1} + \gamma_{a_1}\right)
	\left(i\Delta_{2} + \gamma_{a_2}\right)}
	{\left(i\Delta_{1} + \gamma_{a_1}\right)
	\left(i\Delta_{2} + \gamma_{a_2}\right)
	+ \gamma_1\left(i\Delta_{2} + \gamma_{a_2}\right)
	+ \gamma_2\left(i\Delta_{1} + \gamma_{a_1}\right)
	+ \gamma_1\gamma_2\left(1-e^{i2\theta}\right)}, \\
	S_{22} &= - \frac{e^{i2\theta}\gamma_1\left(i\Delta_{2} + \gamma_{a_2}\right)
	+\gamma_2\left(i\Delta_{1} + \gamma_{a_1}\right)
	+\gamma_1\gamma_2\left(1-e^{i2\theta}\right)}
	{\left(i\Delta_{1} + \gamma_{a_1}\right)
	\left(i\Delta_{2} + \gamma_{a_2}\right)
	+ \gamma_1\left(i\Delta_{2} + \gamma_{a_2}\right)
	+ \gamma_2\left(i\Delta_{1} + \gamma_{a_1}\right)
	+ \gamma_1\gamma_2\left(1-e^{i2\theta}\right)}, \\
\end{align}
\end{widetext}
This result is equivalent to that reported in Refs.\,\onlinecite{Fan2002, Xiao2010}, but with different methods in the derivation. 

\subsection{Simulation results}
\begin{figure*}[ht]
  \centering
  \includegraphics[width=2\columnwidth]{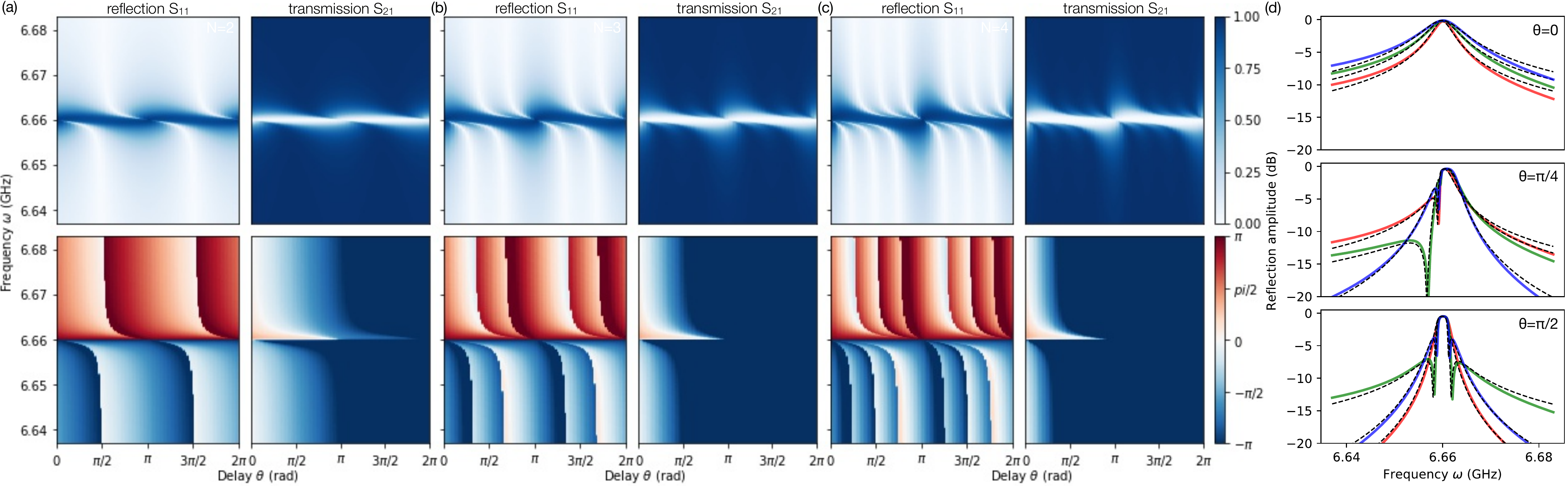}
  \caption{scattering coefficients of a chain of $N$ hanger-type $\lambda/4$ resonators which are coupled to an extended waveguide. (a) For $N=2$, the spectrum periodically changes from a symmetric to an asymmetric line shape with the increase of $\theta$. The symmetric line shapes are obtained at $\theta=n\pi/2$, while there exits a sudden $\pi$ phase change for every $\theta=n\pi+\pi/2$. (b)-(c) With the increase of $N$, the switch becomes more frequently in $\theta$. Here, sudden $\pi$ phase changes occur at $\theta=(2n+1)\pi/2^N$ for $n=1,\cdots,N-1$. (d) The reflection responses for $\theta=0,\pi/4,\pi/2$ (top, middle,bottom) and $N=2,3,4$ (red, green, blue). The black dashed curves indicate the results derived in the system-bath approach. The resonant frequency $\omega_{0}$ is calculated by using the method introduced in Ref.\,\cite{Chen2021a}, which is shifted by $1\,{\rm MHz}$ for a better fitting.}
  \label{fig:simulation_hanger}
\end{figure*}

For large $N$, it is rather tedious to derive an analytical expression of the scattering coefficients and it may be even cumbersome to do numerical simulations. Fortunately, efficient algorithms exist if we simplify the discussion to $\Delta_{j} \equiv \Delta$, $g_{j} \equiv g$, and $\gamma_{a_j} \equiv \gamma_{a}$, as shown in Appendix\,\ref{app:cascade_hanger_numerical}. Here, we perform several numerical simulations for cross checking the results derived above. We choose the real propagating constant $\alpha=5.0\times 10^{-3}\,{\rm /m}$, phase velocity $v_{\rm p}=1.35\times 10^{8}\,{\rm m/s}$, the coupling capacitor $C=1.0\times 10^{-14}\,{\rm F}$, and the resonator length $l=5\times 10^{-3}\,{\rm m}$ for each $\lambda/4$ resonator. The resonant frequency of each individual resonator is estimated to be $\omega_{\rm r}=2\pi\times 6.659\,{\rm GHz}$, with the external and internal decay rates of the resonator, $\gamma=2\pi\times 928\,{\rm kHz}$ and $\gamma_{a}=2\pi\times 212\,{\rm kHz}$ \cite{Chen2021a}. 

In Fig.\,\ref{fig:simulation_hanger}(a), we vary the phase difference, $\theta$, i.e. the distance between neighboring resonators, and calculate the scattering coefficients for $N=2$. On the one hand, the absolute value of the reflection and transmission amplitudes exhibit an asymmetric Fano resonance line shape for $\theta \neq n\pi/2$ with $n=0,1,\cdots$ \cite{Xiao2010}. Depending on whether $\theta=n\pi$ or $n\pi+\pi/2$, we obtain a symmetric Lorentzian spectrum, which is also known as the Breit-Wigner resonance \cite{Breit1936}, or a symmetric Fano spectrum \cite{Miroshnichenko2010}, respectively. These line shapes can be better seen in Fig.\,\ref{fig:simulation_hanger}(d), where we fix $\theta$ to several values. On the other hand, a transition between the symmetric and asymmetric line spectra occurs at $\theta= n\pi/2$. As can bee seen in the corresponding phase diagrams, the transition is smooth at $\theta= n\pi$. However, an abrupt $\pi$ phase shift happens for every $\theta= n\pi+\pi/2$, which distinguishes the symmetric Lorentzian spectrum from the symmetric Fano spectrum . With the increase of $N$, the transition between symmetric and asymmetric line shapes emerge more frequently and occur at $\theta= n\pi+2n'\pi/2^N$ for $n=0,1,\cdots$ and $n'=1,\cdots,n-1$, as shown in Fig.\,\ref{fig:simulation_hanger}(a)-(c). Correspondingly, the sudden $\pi$ phase change happens at $\theta= (2n+1)\pi/2^N$. 

We also compare the reflection line shape for different $N$ and $\theta$ in Fig.\,\ref{fig:simulation_hanger}(d). It can be clearly seen that the phase factor, $\theta$, determines whether a broadening or narrowing of the spectrum is to be observed with an increasing $N$. For example, the full width at half maximum (FWHM) increases moronically with $N$ for $\theta=0$, or equivalently $2\pi$, while it decreases moronically for $\theta=\pi/2$. These observations indicate that, by coupling multiple hanger-type resonators evenly alongside a waveguide, one can engineer the scattering coefficients of the system and obtain a photonic crystal microwave resonator as a whole. Depending on the parameter $\theta$, the new resonator can have a huge enhancement or reduction of the Q factors compared with each individual hanger-type resonator, which provides a new freedom to the resonator design in superconducting quantum circuits.

\section{Coupling multiple necklace-type resonators in a chain} \label{sec:cascade_necklace}
\subsection{General scattering coefficients}
\begin{figure}
\centering
\includegraphics[width=0.9\columnwidth]{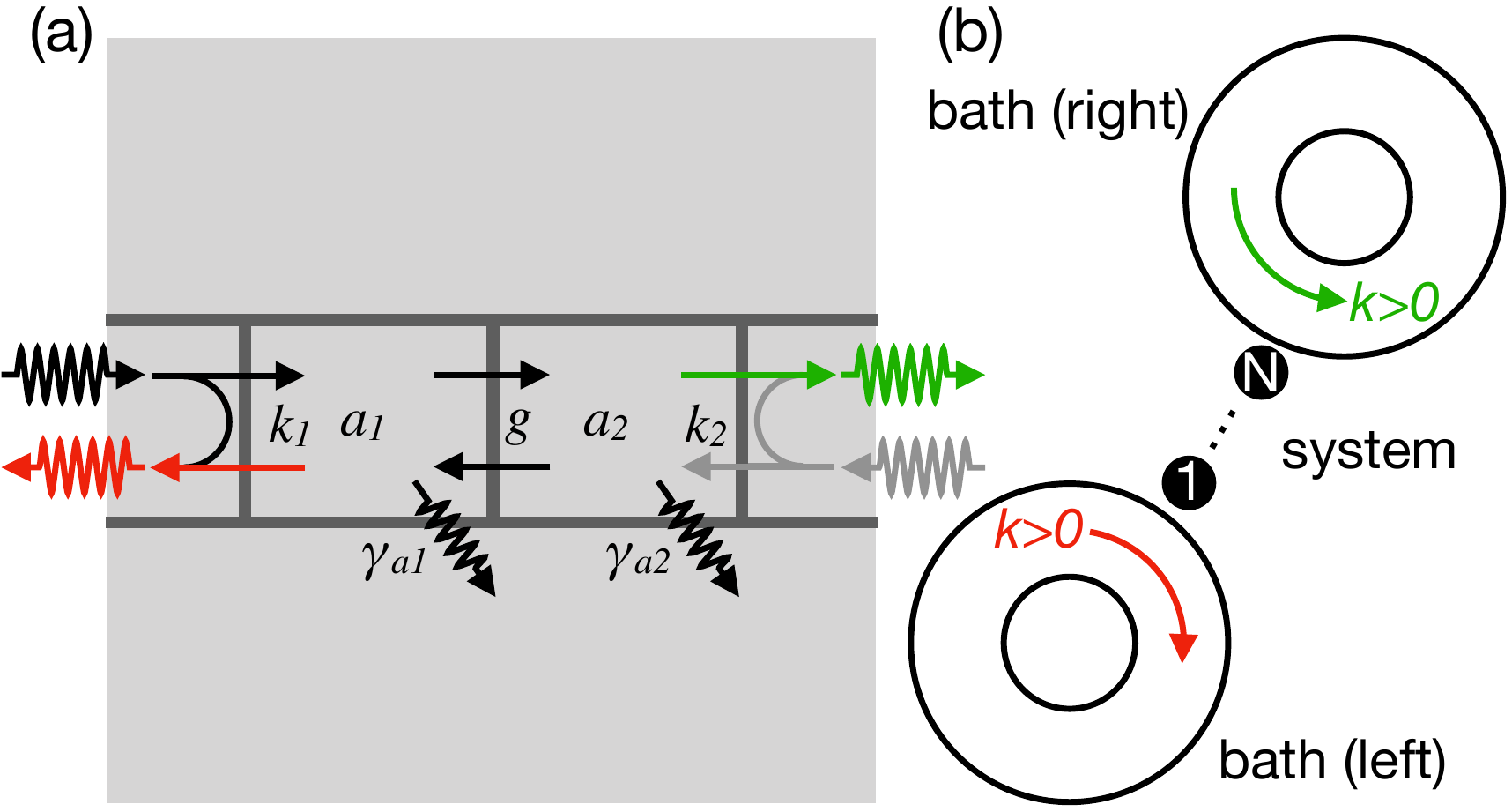}
\caption{Schematic of a chain of coupled necklace-type $\lambda/2$ resonators (i.e., the system), which is coupled to two transmission-line waveguides (i.e., the baths) at the two ends, respectively. Each of the bath accommodates only one unidirectional propagating field. We also use different colors for the outgoing fields emitting to emphasize the $\pi$ phase shift of the intra-resonator spatial mode at the two ends.}
\label{fig:cascade_necklace}
\end{figure}

Let us now consider a system with $N$ necklace-type resonators coupled to each other in a linear chain, as schematically shown in Fig.\,\ref{fig:cascade_necklace}(a)-(b). The total Hamiltonian reads as
\begin{align}
	H_{\rm s} &= \sum_{j=1}^{N}\hbar \Delta_{j} a_{j}^{\dagger}a_{j}
	- \sum_{j=1}^{N-1}\hbar g_{j}\left( a_{j}^{\dagger}a_{j+1}+a_{j}a_{j+1}^{\dagger} \right), \\
	H_{\rm b} &= \sum_{m=1}^{2}\int_{-\infty}^{+\infty}d\omega \hbar \omega b_{m,\omega}^{\dagger} b_{m,\omega}, \\
	H_{\rm sb} &= \int_{-\infty}^{+\infty}d\omega \hbar
	\left(\kappa_{1}^{*} a_1 b_{1,\omega}^{\dagger}  
	+ \kappa_{1}a_1^{\dagger} b_{1,\omega} \right. \nonumber \\
	&\,\,\,\,\,\,\,\,\,\,\,\,\,\,\,\,\,\,\,\,\,\,\,\,\,\,\,\,\,\,
	- \left. \kappa_{2}^{*} a_{N} b_{2,\omega}^{\dagger}  
	- \kappa_{2}a_{N}^{\dagger} b_{2,\omega} \right). 
\end{align}
The input-output relations can be readily obtained by following the standard procedure
\begin{align}
	b_{1,{\rm out}} = b_{1,{\rm in}} + \sqrt{\gamma_{1}}a_{1}, \label{eq:cascade_necklace_in_out_b1} \\
	b_{2,{\rm out}} = b_{2,{\rm in}} - \sqrt{\gamma_{2}}a_{N}, \label{eq:cascade_necklace_in_out_b2}
\end{align}
and, in turn, the evolution of the intra-resonator reads
\begin{widetext}
\begin{align}
	\dot{a}_j = \begin{cases}
  -i\Delta_{1}a_1 
	+ ig_{1}a_{2}
	- \left(\frac{\gamma_1}{2}+\frac{\gamma_{a_1}}{2}\right)a_{1}
	- \sqrt{\gamma_{1}}b_{1,{\rm in}}, &\text{for}\,j=1; \\
  -i\Delta_{j}a_j 
	+ i\left(g_{j-1}a_{j-1} + g_{j}a_{j+1}\right)
	-\frac{\gamma_{a_j}}{2}a_{j}, &\text{for}\,1<j<N; \\
  -i\Delta_{N}a_N 
	+ ig_{N-1}a_{N-1}
	- \left(\frac{\gamma_2}{2}+\frac{\gamma_{a_N}}{2}\right)a_{N}
	+ \sqrt{\gamma_{2}}b_{2,{\rm in}}, &\text{for}\,j=N.
\end{cases}
\label{eq:cascade_necklace_in_out_a}
\end{align}
\end{widetext}

Similar to the single-resonator case, the scattering coefficients can be obtained by finding the mean-field steady-state solution of Eqs.\,\eqref{eq:cascade_necklace_in_out_b1}-\eqref{eq:cascade_necklace_in_out_a}. For example, for the simplest case with $N=2$, we have
\begin{align}
	S_{11} &= 1-\frac{\gamma_{1} \left[i\Delta_{2} + \left(\frac{\gamma_{2}}{2}+\frac{\gamma_{a_2}}{2}\right)\right]}{\left[i\Delta_{1} + \left(\frac{\gamma_{1}}{2}+\frac{\gamma_{a_1}}{2}\right) \right]\left[i\Delta_{2} + \left(\frac{\gamma_{2}}{2}+\frac{\gamma_{a_2}}{2}\right) \right] + g^2}, \\
	S_{21} &= S_{12} = \frac{-ig\sqrt{\gamma_{1}\gamma_{2}}}{\left[i\Delta_{1} + \left(\frac{\gamma_{1}}{2}+\frac{\gamma_{a_1}}{2}\right) \right]\left[i\Delta_{2} + \left(\frac{\gamma_{2}}{2}+\frac{\gamma_{a_2}}{2}\right) \right] + g^2}, \\
	S_{22} &= 1-\frac{\gamma_{2} \left[i\Delta_{1} + \left(\frac{\gamma_{1}}{2}+\frac{\gamma_{a_1}}{2}\right)\right]}{\left[i\Delta_{1} + \left(\frac{\gamma_{1}}{2}+\frac{\gamma_{a_1}}{2}\right) \right]\left[i\Delta_{2} + \left(\frac{\gamma_{2}}{2}+\frac{\gamma_{a_2}}{2}\right) \right] + g^2}.
\end{align}
This result has been derived and experimentally demonstrated in our previous work \cite{Fischer2021}. 

\subsection{The tight-binding model}
To gain further physical insight beyond the numerical solutions for large $N$, we further assume that $\Delta_{j} \equiv \Delta$, $g_{j} \equiv g$, and $\gamma_{a_j} \equiv \gamma_{a}$, such that the resonator chain can be described by a bosonic tight-binding model. For a finite number of resonators, $N$, we define the following collective modes of the resonator chain 
\begin{align}
	a_{k} = \sum_{j=1}^{N}\sqrt{\frac{2}{N+1}}\sin\left(\frac{\pi kj}{N+1}\right) a_{j}. 
	\label{eq:cascade_necklace_magnon}
\end{align}
This replacement of variable is similar to the definition of magnon in spin systems \cite{Wojcik2005, *Wojcik2007}. In this way, we can rewrite Eqs.\,\eqref{eq:cascade_necklace_in_out_b1}-\eqref{eq:cascade_necklace_in_out_a} in a more compact form
\begin{align}
	l_{\rm out} &= l_{\rm in} + \sum_{k=1}^{N}\sqrt{\gamma_{1,k}}a_{k}, \\
	r_{\rm out} &= r_{\rm in} + \sum_{k=1}^{N}(-1)^{k}\sqrt{\gamma_{2,k}}a_{k},\\
	\dot{a}_{k} &= -\left(i\Delta_k + \frac{\gamma_{a}}{2}\right)a_{k} 
	- \left[\sqrt{\gamma_{1,k}}b_{1,{\rm in}} + (-1)^{k}\sqrt{\gamma_{2,k}}b_{2,{\rm in}}\right] \nonumber \\
	&- \frac{1}{2}\sum_{k'=1}^{N}\left[\sqrt{\gamma_{1,k}\gamma_{1,k'}}
	+(-1)^{k+k'}\sqrt{\gamma_{2,k}\gamma_{2,k'}}\right] a_{k'}. \label{eq:cascade_necklace_tight_binding_a}
\end{align}
Here, $\Delta_k=\Delta - 2g\cos\left(\frac{k\pi}{N+1}\right)$, $\sqrt{\gamma_{m,k}}=\sqrt{2\gamma_m/(N+1)}\sin\left[k\pi/(N+1)\right]$. With these results, one can readily obtain the following analytical expressions for the scattering coefficients (see Appendix\,\ref{app:cascade_necklace} for detail)
\begin{widetext}
\begin{align}
	S_{11} &= 1-\frac{\left[\left(1+\frac{1}{2}\sum_{k=1}^{N}\frac{\gamma_{2,k}}{i\Delta_k+\frac{\gamma_a}{2}}\right)
	\left(\sum_{k=1}^{N}\frac{\gamma_{1,k}}{i\Delta_k+\frac{\gamma_a}{2}}\right)
	-\frac{1}{2}\left(\sum_{k=1}^{N}\frac{(-1)^{k}\sqrt{\gamma_{1,k}\gamma_{2,k}}}{i\Delta_{k}+\frac{\gamma_{a}}{2}}\right)^2\right]}
	{\left(1+\frac{1}{2}\sum_{k=1}^{N}\frac{\gamma_{1,k}}{i\Delta_{k}+\frac{\gamma_{a}}{2}}\right)
	\left(1+\frac{1}{2}\sum_{k=1}^{N}\frac{\gamma_{2,k}}{i\Delta_{k}+\frac{\gamma_{a}}{2}}\right)
	-\left(\frac{1}{2}\sum_{k=1}^{N}\frac{(-1)^{k}\sqrt{\gamma_{1,k}\gamma_{2,k}}}{i\Delta_{k}+\frac{\gamma_{a}}{2}}\right)^2},\\
	S_{21} &= S_{12} = 
	-\frac{\left(\sum_{k=1}^{N}\frac{(-1)^{k}\sqrt{\gamma_{1,k}\gamma_{2,k}}}{i\Delta_k+\frac{\gamma_a}{2}}\right)}
	{\left(1+\frac{1}{2}\sum_{k=1}^{N}\frac{\gamma_{1,k}}{i\Delta_{k}+\frac{\gamma_{a}}{2}}\right)
	\left(1+\frac{1}{2}\sum_{k=1}^{N}\frac{\gamma_{2,k}}{i\Delta_{k}+\frac{\gamma_{a}}{2}}\right)
	-\left(\frac{1}{2}\sum_{k=1}^{N}\frac{(-1)^{k}\sqrt{\gamma_{1,k}\gamma_{2,k}}}{i\Delta_{k}+\frac{\gamma_{a}}{2}}\right)^2},\\
	S_{22} &= 1-\frac{\left[\left(1+\frac{1}{2}\sum_{k=1}^{N}\frac{\gamma_{1,k}}{i\Delta_k+\frac{\gamma_a}{2}}\right)
	\left(\sum_{k=1}^{N}\frac{\gamma_{2,k}}{i\Delta_k+\frac{\gamma_a}{2}}\right)
	-\frac{1}{2}\left(\sum_{k=1}^{N}\frac{(-1)^{k}\sqrt{\gamma_{1,k}\gamma_{2,k}}}{i\Delta_{k}+\frac{\gamma_{a}}{2}}\right)^2\right]}
	{\left(1+\frac{1}{2}\sum_{k=1}^{N}\frac{\gamma_{1,k}}{i\Delta_{k}+\frac{\gamma_{a}}{2}}\right)
	\left(1+\frac{1}{2}\sum_{k=1}^{N}\frac{\gamma_{2,k}}{i\Delta_{k}+\frac{\gamma_{a}}{2}}\right)
	-\left(\frac{1}{2}\sum_{k=1}^{N}\frac{(-1)^{k}\sqrt{\gamma_{1,k}\gamma_{2,k}}}{i\Delta_{k}+\frac{\gamma_{a}}{2}}\right)^2}.
\end{align}	
\end{widetext}

We note that the replacement of variables in Eq.\,\eqref{eq:cascade_necklace_magnon} is valid for a linear chain of a finite number of microwave resonators. However, one may also assume a periodic boundary condition which describes a $N$-resonator loop or an infinitly long chain of resonators with $N\rightarrow\infty$
\begin{align}
	a_{k} = \sum_{j=1}^{N}\sqrt{\frac{1}{N}}\exp\left(\frac{i2\pi kj}{N}\right) a_{j}.
\end{align}
This replacement leads to a slightly different dispersive relation $\Delta_k=\Delta - 2g\cos\left(2k\pi/N\right)$, and the damping rates $\sqrt{\gamma_{1,k}}=\sqrt{\gamma_1/N}\exp\left[-i2\pi k/N\right]$, $\sqrt{\gamma_{2,k}}=\sqrt{\gamma_2/N}$. The input-output relations are 
\begin{align}
	l_{\rm out} &= l_{\rm in} + \sum_{k=1}^{N}\sqrt{\gamma_{1,k}}a_{k}, \\
	r_{\rm out} &= r_{\rm in} - \sum_{k=1}^{N}\sqrt{\gamma_{2,k}}a_{k},\\
	\dot{a}_{k} &= -\left(i\Delta_k + \frac{\gamma_{a}}{2}\right)a_{k} 
	- \left[\sqrt{\gamma_{1,k}}b_{1,{\rm in}}-\sqrt{\gamma_{2,k}}b_{2,{\rm in}}\right] \nonumber \\
	&- \frac{1}{2}\sum_{k'=1}^{N}\left[\sqrt{\gamma_{1,k}\gamma_{1,k'}}
	+\sqrt{\gamma_{2,k}\gamma_{2,k'}}\right] a_{k'}, \label{eq:cascade_necklace_tight_binding_a}
\end{align}
and the scattering parameters can be expressed as
\begin{widetext}
\begin{align}
	S_{11} &= 1-\frac{\left[\left(1+\frac{1}{2}\sum_{k=1}^{N}\frac{\gamma_{2,k}}{i\Delta_k+\frac{\gamma_a}{2}}\right)
	\left(\sum_{k=1}^{N}\frac{\gamma_{1,k}}{i\Delta_k+\frac{\gamma_a}{2}}\right)
	-\frac{1}{2}\left(\sum_{k=1}^{N}\frac{\sqrt{\gamma_{1,k}\gamma_{2,k}}}{i\Delta_{k}+\frac{\gamma_{a}}{2}}\right)^2\right]}
	{\left(1+\frac{1}{2}\sum_{k=1}^{N}\frac{\gamma_{1,k}}{i\Delta_{k}+\frac{\gamma_{a}}{2}}\right)
	\left(1+\frac{1}{2}\sum_{k=1}^{N}\frac{\gamma_{2,k}}{i\Delta_{k}+\frac{\gamma_{a}}{2}}\right)
	-\left(\frac{1}{2}\sum_{k=1}^{N}\frac{\sqrt{\gamma_{1,k}\gamma_{2,k}}}{i\Delta_{k}+\frac{\gamma_{a}}{2}}\right)^2},\\
	S_{21} &= S_{12} = 
	\frac{\left(\sum_{k=1}^{N}\frac{\sqrt{\gamma_{1,k}\gamma_{2,k}}}{i\Delta_k+\frac{\gamma_a}{2}}\right)}
	{\left(1+\frac{1}{2}\sum_{k=1}^{N}\frac{\gamma_{1,k}}{i\Delta_{k}+\frac{\gamma_{a}}{2}}\right)
	\left(1+\frac{1}{2}\sum_{k=1}^{N}\frac{\gamma_{2,k}}{i\Delta_{k}+\frac{\gamma_{a}}{2}}\right)
	-\left(\frac{1}{2}\sum_{k=1}^{N}\frac{\sqrt{\gamma_{1,k}\gamma_{2,k}}}{i\Delta_{k}+\frac{\gamma_{a}}{2}}\right)^2},\\
	S_{22} &= 1-\frac{\left[\left(1+\frac{1}{2}\sum_{k=1}^{N}\frac{\gamma_{1,k}}{i\Delta_k+\frac{\gamma_a}{2}}\right)
	\left(\sum_{k=1}^{N}\frac{\gamma_{2,k}}{i\Delta_k+\frac{\gamma_a}{2}}\right)
	-\frac{1}{2}\left(\sum_{k=1}^{N}\frac{\sqrt{\gamma_{1,k}\gamma_{2,k}}}{i\Delta_{k}+\frac{\gamma_{a}}{2}}\right)^2\right]}
	{\left(1+\frac{1}{2}\sum_{k=1}^{N}\frac{\gamma_{1,k}}{i\Delta_{k}+\frac{\gamma_{a}}{2}}\right)
	\left(1+\frac{1}{2}\sum_{k=1}^{N}\frac{\gamma_{2,k}}{i\Delta_{k}+\frac{\gamma_{a}}{2}}\right)
	-\left(\frac{1}{2}\sum_{k=1}^{N}\frac{\sqrt{\gamma_{1,k}\gamma_{2,k}}}{i\Delta_{k}+\frac{\gamma_{a}}{2}}\right)^2}.
\end{align}	
\end{widetext}

\subsection{Simulation results}
\begin{figure*}
  \centering
  \includegraphics[width=2\columnwidth]{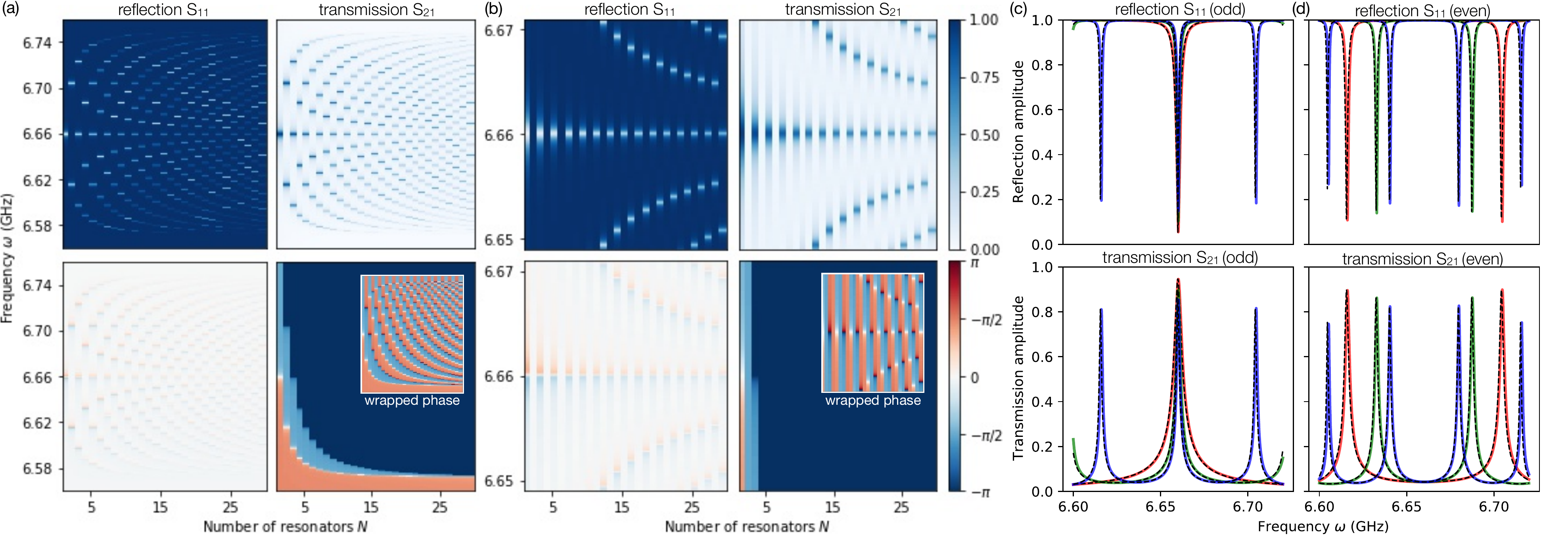}
  \caption{scattering coefficients of a chain of coupled necklace-type $\lambda/4$ resonators, which couple to two feedlines at the ends. (a) With the increase of the resonator number, $N$, there emerges the same number of resonance dips or peak in the reflection and transmission responses. The spread of the peaks (dips) saturates at a value which equals to four times of the coupling strength. (b) A closer inspection of the spectrum around the central resonant frequency indicate that the incident photons can be almost fully reflected or transmitted depending on the parity of $N$. The FWHM of the central peak also decreases moronically with $N$. (c) The reflection and transmission responses for $N=2,3,4$ (red, green, blue), respectively. The black dashed curves indicate the results derived in the system-bath approach. The resonant frequency $\omega_{0}$ is calculated by using the method introduced in Ref.\,\cite{Chen2021a}, which is shifted by $1\,{\rm MHz}$ for a better fitting.}
  \label{fig:simulation_necklace}
\end{figure*}

As a crosscheck of the results derived above, we performed several numerical simulations with the same parameters as before. However, the resonator length should be extended to $l=10\times 10^{-3}\,{\rm m}$ for $\lambda/2$ resonators. For each individual resonator, the resonant frequency is estimated to be $\omega_{\rm r}=2\pi\times 6.659\,{\rm GHz}$. The decay rates are $\gamma=2\pi\times 1.86\,{\rm MHz}$ and $\gamma_{a}=2\pi\times 212\,{\rm kHz}$. 

In Fig.\,\ref{fig:simulation_hanger}(a), we calculate the scattering coefficients for increasing $N$. On the one hand, we observe that the number of resonant frequencies in the spectrum increases with the number of resonators. On the other hand, the spread of the resonant frequencies of the whole system also scales with $N$ but saturates to a finite value. The bandwidth is approximately four times of the coupling strength, $g \approx 2\pi\times 44\,{\rm MHz}$, which is calculated by using the method introduced in Ref.\,\cite{Chen2021a}. This phenomena can be readily understood in the dispersion relation of the collective fields $a_{k}$. A more detailed inspection of the spectrum around $\omega_{\rm r}$ indicates that the photons at the central resonant frequency can be almost fully transmitted or reflected, depending on whether there is an \textit{odd} or \textit{even} number of resonators in the chain, as shown in Fig.\,\ref{fig:simulation_hanger}(b). This observation indicates a possible application that a chain of coupled resonators can be used as a single-photon switch \cite{Liao2009, Liao2010}. Moreover, we observe that the FWHM of the line shape decreases monotonically with $N$, as shown in Figs.\,\ref{fig:simulation_hanger}(c). It indicates the existence of high Q-factor mode in an array of coupled resonators, which has been theoretically predicted by using the transfer-matrix method and experimentally demonstrated in the literature \cite{Sumetsky2003, Hara2005, OBrien2007, Notomi2008}. Together with the result of a chain of hanger-type resonators, we conclude that coupling multiple microwave resonators can fundamentally change the scattering coefficients of the system and provide a novel way to increase the Q factors of superconducting microwave resonators.

\section{Conclusions and outlook} \label{sec:conclusions}
In conclusion, we provide a systematic study of the scattering coefficients of superconducting microwave resonators in the quantum perspective. By transforming the Hamiltonian from the wave vector space to the frequency space, we describe a unified approach for input-output analysis that applies to a general microwave resonator. The corresponding scattering coefficients are consistent with those derived in the classical transfer-matrix approach \cite{Chen2021a}. We also generalize our method to more complex systems with multiple hanger- or necklace-type resonators and time delays. We reveal several interesting photon transport phenomena in such photonic-crystal-like systems and find consistency with the results reported in the literature.

Compared with the transfer-matrix approach, the system-bath method has the advantage of simplicity which requires no detailed information of the circuit parameters. In this regard, it is more suitable to use the quantum approach to analyze the scattering coefficients of a complex circuit. However, the classical approach is more close to a physical system, which directly relates the electrical properties of a resonator, such as resonant frequency and quality factors, to the circuit parameters such as the coupling capacitance and the characteristic impedance. Nevertheless, a full description of a network of superconducting microwave resonators can be obtained by using either of the two approaches. They together provide a firm ground for the study of transport properties in a general superconducting quantum circuit.

As a closing remark, we note that dephasing of microwave resonators can also be incorporated in this theoretical frame, which is, however, often neglected in the literature. To describe dephasing, we write the system-bath interaction in the following form \cite{Walls1985, Gardiner2000, Turchette2000, Myatt2000, Liu2004}
\begin{align}
	H_{\rm sb} &= \sum_{m=1}^{2}\left(-1\right)^{m-1}\int_{-\infty}^{+\infty}d\omega \hbar
	a^{\dagger} a \left(\kappa_{\phi}^{*} b_{m,\omega}^{\dagger}  
	+ \kappa_{\phi} b_{m,\omega}  \right).
\end{align}
The resulting input-output relation can be readily derived by following the same procedure as described above. The result is
\begin{align}
	\dot{a} &= -i\Delta_{a}a 
	- \frac{\gamma_{\phi}}{2} a 
	- \sqrt{\gamma_{\phi}}\left(ab_{\rm in}-b_{\rm in}^{\dagger}a\right), \label{eq:langevin_dephasing}
\end{align}
where $\sqrt{\gamma_{\phi}} = i\sqrt{2\pi}\kappa_{\phi}$. We note that the input field, $b_{\rm in}=\int_{-\infty}^{+\infty} d\omega e^{+i\omega t}b_{\omega}(0)/\sqrt{2\pi}$, does not commute with the intra-resonator field. However, the combined operator $b_{\rm in} + \left(\sqrt{\gamma_\phi}\right)^{*}a^{\dagger}a/2$ and $b_{\rm in}^{\dagger} + \sqrt{\gamma_{\phi}}a^{\dagger}a/2$ commute with all the system operators \cite{Gardiner2000}, such that we rewrite Eq.\,\eqref{eq:langevin_dephasing} as
\begin{align}
	\dot{a} &= -i\Delta_{a}a 
	- \sqrt{\gamma_{\phi}}a\left(b_{\rm in}-b_{\rm in}^{\dagger}\right).
\end{align}
This result indicates that the input field $b$ causes a random jittering in the resonant frequency of the intra-resonator field, $a$. Even though it conserves the energy of the system, the dephasing effect broadens the line shape of the scattering responses such that one may not simply extract the energy dissipation rate from the measured FWHM of scattering spectrum. In this regard, one should resort to an extra measurement, for example, Ramsey interferometry, to distinguish the contributions of energy decay and dephasing in the FWHM. Today, high quality microwave resonators with $Q_{\rm i}>10^{6}$ can be routinely made in lab \cite{Megrant2012}. We anticipate that a careful calibration and analysis of the dephasing rate may pave a significant step towards making even higher-quality superconducting microwave resonators in the near future.

\begin{acknowledgments}
We acknowledge support by German Research Foundation via Germany's Excellence Strategy (EXC-2111-390814868), Elite Network of Bavaria through the program ExQM, European Union via the Quantum Flagship project QMiCS (No.\,820505), German Federal Ministry of Education and Research via the project QuaRaTe (No.\,13N15380).
\end{acknowledgments}

\appendix
\begin{widetext}
\section{Outline of the input-output analysis} \label{app:input_output} 
We model a microwave resonator of any type as a composite system which the following Hamiltonian \cite{Gardiner1985, Gardiner1993}
\begin{align}
	H_{\rm s} &= \sum_{n}\hbar \omega_{n}a_{n}^{\dagger}a_{n}, \label{eq:langevin_hamiltonian_system} \\
	H_{\rm b} &= \sum_{m}\hbar\int_{-\infty}^{+\infty} d\omega \omega b_{m,\omega}^{\dagger}b_{m,\omega}, \label{eq:langevin_hamiltonian_bath} \\
	H_{\rm sb} &= \sum_{m,n}\hbar\int_{-\infty}^{+\infty} d\omega \left[\kappa_{m,n}^{*}a_{n}b_{m,\omega}^{\dagger} 
	+ \kappa_{m,n}a_{n}^{\dagger}b_{m,\omega} \right]. \label{eq:langevin_hamiltonian_interaction}
\end{align}
Following the Heisenberg equation, the time evolution of the intra-resonator field, $a$, as well as the bath field, $b_{\omega}$, can be readily written as 
\begin{align}
	\dot{a}_{n} &= -i\omega_{n}a_{n}
	-i\sum_{m}\int_{-\infty}^{+\infty}d\omega \kappa_{m,n} l_{m}, \label{eq:langevin_heisenberg_a}\\
	\dot{l}_{m} &= -i\omega l_{m} - i\sum_{n}\kappa_{m,n}^{*}a_{n}. \label{eq:langevin_heisenberg_b}
\end{align}
By inserting the formal solution of Eq.\,\eqref{eq:langevin_heisenberg_b},
\begin{align}
	b_{m,\omega}(t) &= e^{-i\omega (t-t_0)} b_{m,\omega}(t_0) - i\sum_{n}\kappa_{m,n'}^{*}
	\int_{t_0}^{t} dt' e^{-i\omega (t-t')} a_{n'}(t').
\end{align}
into Eq.\,\eqref{eq:langevin_heisenberg_a}, we obtain
\begin{align}
	\dot{a}_{n} &= -i\omega_{n}a_{n}
	-i\sum_{m}\kappa_{m,n} \int_{-\infty}^{+\infty}d\omega e^{-i\omega (t-t_0)} b_{m,\omega}(t_0) 
	-\sum_{m,n'}\kappa_{m,n}\kappa_{m,n'}^{*} \int_{-\infty}^{+\infty}d\omega
	\int_{t_0}^{t} dt' e^{-i\omega (t-t')} a_{n'}(t')
\end{align}

We recall the property 
\begin{align}
	\int_{-\infty}^{+\infty}d\omega e^{-i\omega\left(t-t_0\right)} = 2\pi\delta\left(t-t_0\right)
\end{align}
and defined the input field, i.e., the noise operator in quantum Langevin equation, as
\begin{align}
	b_{m,{\rm in}} = \frac{1}{\sqrt{2\pi}}\int_{-\infty}^{+\infty}d\omega e^{-i\omega (t-t_0)} b_{m,\omega}(t_0), \label{eq:langevin_input_b}\end{align}
the above equation can be simplified as
\begin{align}
	\dot{a}_{n} &= -i\omega_{n}a_{n}
	-i\sum_{m}\sqrt{2\pi}\kappa_{m,n}b_{m,{\rm in}}
	-\sum_{m,n'}\pi\kappa_{m,n}\kappa_{m,n'}^{*} a_{n'}. \label{eq:langevin_input_a}
\end{align}

Alternatively, the formal solution of Eq.\,\eqref{eq:langevin_heisenberg_b} may be written as 
\begin{align}
	b_{m,\omega}(t) &= e^{-i\omega (t-t_1)} b_{m,\omega}(t_1) + i\sum_{n}\kappa_{m,n'}^{*}
	\int_{t}^{t_1} dt' e^{-i\omega (t-t')} a_{n'}(t').
\end{align}
We have
\begin{align}
	\dot{a}_{n} &= -i\omega_{n}a_{n}
	-i\sum_{m}\kappa_{m,n} \int_{-\infty}^{+\infty}d\omega e^{-i\omega (t-t_1)} b_{m,\omega}(t_1) 
	+\sum_{m,n'}\kappa_{m,n}\kappa_{m,n'}^{*} \int_{-\infty}^{+\infty}d\omega
	\int_{t}^{t_1} dt' e^{-i\omega (t-t')} a_{n'}
\end{align}
We defined the output field as
\begin{align}
	b_{m,{\rm out}} = \frac{1}{\sqrt{2\pi}}\int_{-\infty}^{+\infty}d\omega e^{-i\omega (t-t_1)} b_{m,\omega}(t_0), \label{eq:langevin_output_b}
\end{align}
such that the above equation can be simplified as
\begin{align}
	\dot{a}_{n} &= -i\omega_{n}a_{n}
	-i\sum_{m}\sqrt{2\pi}\kappa_{m,n}b_{m,{\rm out}}
	+\sum_{m,n'}\pi\kappa_{m,n}\kappa_{m,n'}^{*} a_{n'}. \label{eq:langevin_output_a}
\end{align}

Combining Eqs.\,\eqref{eq:langevin_input_b}-\eqref{eq:langevin_input_a} and \eqref{eq:langevin_output_b}-\eqref{eq:langevin_output_a}, we obtain the so-called input-output analysis
\begin{align}
	\dot{a}_{n} &= -i\omega_{n}a_{n}
	-\sum_{m}\sqrt{\gamma_{m,n}}b_{m,{\rm in}}
	-\frac{1}{2}\sum_{m,n'}\sqrt{\gamma_{m,n}}\left(\sqrt{\gamma_{m,n'}}\right)^{*} a_{n'}
	-\frac{\gamma_{a_n}}{2}a_{n}, \label{eq:inout_a_app}\\
	b_{m,{\rm out}} &= b_{m,{\rm in}} + \sum_{n'} \left(\sqrt{\gamma_{m,n'}}\right)^{*}a_{n'}. , \label{eq:inout_b_app}
\end{align}
Here, we have defined $\sqrt{\gamma_{m,n}} = i\sqrt{2\pi}\kappa_{m,n}$ and $\left(\sqrt{\gamma_{m,n}}\right)^{*} = -i\sqrt{2\pi}\kappa_{m,n}^{*}$ and add the intrinsic damping of the oscillator by hand.

\section{A chain of hanger-type resonators with delays}\label{app:cascade_hanger}
We consider a composite system where $N$ hanger-type resonators are side-coupled to a 1D waveguide. The system-bath interaction reads
\begin{align}
	H_{\rm sb} &= \sum_{j=1}^{N}\int_{-\infty}^{+\infty}d\omega \hbar
	\left\{ e^{-i(j-1)\omega\tau}\kappa_{j}^{*} a_{j} l_{\omega}^{\dagger}
	+ e^{i(j-1)\omega\tau}\kappa_{j} a_{j}^{\dagger} l_{\omega} 
	+ e^{-i(j-1)\omega\tau}\kappa_{j}^{*} a_{j} r_{\omega}^{\dagger}
	+ e^{i(j-1)\omega\tau}\kappa_{j} a_{j}^{\dagger} r_{\omega}\right\}.
\end{align}
Here, we encode the information of the distance between different hanger-type resonators into a phase delay $l_{\omega} \rightarrow e^{i(j-1)\omega\tau}l_{\omega}$, $r_{\omega} \rightarrow e^{i(j-1)\omega\tau}r_{\omega}$, $j=1,\cdots,N$. Following the same procedure in Appendix\,\ref{app:input_output}, we describe the dynamics of the intra-resonator field, $a_{j}$, and the two bath fields, $l_{\omega}$ and $r_{\omega}$, as
\begin{align}
	l_{\omega}(t) &= e^{+i\omega (t-t_0)} l_{\omega}(t_0) - i\sum_{n}e^{-i(j-1)\omega\tau}\kappa_{m,n'}^{*}
	\int_{t_0}^{t} dt' e^{+i\omega (t-t')} a_{n'}(t'), \\
	r_{\omega}(t) &= e^{-i\omega (t-t_0)} r_{\omega}(t_0) - i\sum_{n}e^{-i(j-1)\omega\tau}\kappa_{m,n'}^{*}
	\int_{t_0}^{t} dt' e^{-i\omega (t-t')} a_{n'}(t'). \\
	\dot{a}_{j} &= -i\omega_{j}a_{j}
	-i \kappa_{j} \int_{-\infty}^{+\infty}d\omega e^{+i\omega (t-t_0)}e^{i(j-1)\omega\tau} l_{\omega}(t_0)  
	-\sum_{j'}\kappa_{j}\kappa_{j'}^{*} \int_{-\infty}^{+\infty}d\omega
	\int_{t_0}^{t} dt' e^{+i\omega (t-t')}e^{i(j-j')\omega\tau} a_{n'}(t') \nonumber \\
	&-i \kappa_{j} \int_{-\infty}^{+\infty}d\omega e^{-i\omega (t-t_0)}e^{-i(j-1)\omega\tau} r_{\omega}(t_0) 
	-\sum_{j'}\kappa_{j}\kappa_{j'}^{*} \int_{-\infty}^{+\infty}d\omega
	\int_{t_0}^{t} dt' e^{-i\omega (t-t')}e^{i(j-j')\omega\tau} a_{n'}(t') \label{eq:cascade_hanger_in_a1}
\end{align}
We define the input fields $l_{\rm in}=\left(1/\sqrt{2\pi}\right)\int_{-\infty}^{+\infty} e^{+i\omega (t-t_0)} l_{\omega}(t_0)d\omega$, $r_{\rm in}=\left(1/\sqrt{2\pi}\right)\int_{-\infty}^{+\infty} e^{+i\omega (t-t_0)} r_{\omega}(t_0)d\omega$, such that Eq.\,\eqref{eq:cascade_hanger_in_a1} can be written in a compact form  
\begin{align}
	\dot{a}_{j} &= -i\omega_{j}a_{j} 
	- 2\pi \kappa_{j}\kappa_{j}^{*} a_{j}
	-i \sqrt{2\pi}\kappa_{j} l_{\rm in}\left(t+(j-1)\tau\right)  
	-i \sqrt{2\pi}\kappa_{j} r_{\rm in}\left(t-(j-1)\tau\right) \nonumber \\
	&-2\pi\sum_{j'>j}\kappa_{j}\kappa_{j'}^{*} a_{j'}\left[t+(j-j')\tau\right]
	-2\pi\sum_{j'<j}\kappa_{j}\kappa_{j'}^{*} a_{j'}\left[t+(j'-j)\tau\right] \label{eq:cascade_hanger_in_a2}
\end{align}
Here, a technical problem emerges that the time evolution of $a_j$ involves operators at different times. To eliminate the time dependance of $\tau$, we note that the above discussion is performed in the rotating frame at the driving frequency $\omega_{\rm d}$. Thus, a time delay of $\tau$ in the operators may be fairly approximated by a phase factor $\exp\left(i\theta\right)$ with $\theta = \omega_{\rm d} \tau$. In this regard, we rewrite Eq.\,\eqref{eq:cascade_hanger_in_a2} as 
\begin{align}
	\dot{a}_{j} &= -i\omega_{j}a_{j} 
	- 2\pi \kappa_{j}\kappa_{j}^{*} a_{j}
	-i \sqrt{2\pi}\kappa_{j} l_{\rm in}\left(t\right)e^{-i(j-1)\theta}
	-i \sqrt{2\pi}\kappa_{j} r_{\rm in}\left(t\right)e^{i(j-1)\theta} \nonumber \\
	&-2\pi\sum_{j'>j}\kappa_{j}\kappa_{j'}^{*}e^{(j'-j)\theta} a_{j'}(t)
	-2\pi\sum_{j'<j}\kappa_{j}\kappa_{j'}^{*}e^{(j-j')\theta}a_{j'}(t) \label{eq:cascade_hanger_in_a3}
\end{align}

On the other hand, we have
\begin{align}
    l_{\omega}(t) &= e^{+i\omega (t-t_1)} l_{\omega}(t_1) 
    + i\sum_{n}e^{-i(j-1)\omega\tau}\kappa_{m,n'}^{*}
	\int_{t}^{t_1} dt' e^{+i\omega (t-t')} a_{n'}(t'), \\
	r_{\omega}(t) &= e^{-i\omega (t-t_1)} r_{\omega}(t_1) 
	+ i\sum_{n}e^{-i(j-1)\omega\tau}\kappa_{m,n'}^{*}
	\int_{t}^{t_1} dt' e^{-i\omega (t-t')} a_{n'}(t'). \\
	\dot{a}_{j} &= -i\omega_{j}a_{j}
	-i \kappa_{j} \int_{-\infty}^{+\infty}d\omega e^{+i\omega (t-t_1)}e^{i(j-1)\omega\tau} l_{\omega}(t_1)  
	+\sum_{j'}\kappa_{j}\kappa_{j'}^{*} \int_{-\infty}^{+\infty}d\omega
	\int_{t}^{t_1} dt' e^{+i\omega (t-t')}e^{i(j-j')\omega\tau} a_{n'}(t') \nonumber \\
	&-i \kappa_{j} \int_{-\infty}^{+\infty}d\omega e^{-i\omega (t-t_1)}e^{-i(j-1)\omega\tau} r_{\omega}(t_1) 
	+\sum_{j'}\kappa_{j}\kappa_{j'}^{*} \int_{-\infty}^{+\infty}d\omega
	\int_{t}^{t_1} dt' e^{-i\omega (t-t')}e^{i(j-j')\omega\tau} a_{n'}(t')
\end{align}
We define the output fields $l_{\rm out}=\left(1/\sqrt{2\pi}\right)\int_{-\infty}^{+\infty} e^{+i\omega (t-t_1)} l_{\omega}(t_1)d\omega$, $r_{\rm out}=\left(1/\sqrt{2\pi}\right)\int_{-\infty}^{+\infty} e^{+i\omega (t-t_1)} r_{\omega}(t_1)d\omega$, and obtain
\begin{align}
	\dot{a}_{j} &= -i\omega_{j}a_{j} 
	+ 2\pi \kappa_{j}\kappa_{j}^{*} a_{j}
	-i \sqrt{2\pi}\kappa_{j} l_{\rm out}\left(t\right)e^{-i(j-1)\theta}
	-i \sqrt{2\pi}\kappa_{j} r_{\rm out}\left(t\right)e^{i(j-1)\theta} \nonumber \\
	&+2\pi\sum_{j'<j}\kappa_{j}\kappa_{j'}^{*}e^{(j'-j)\theta} a_{j'}(t)
	+2\pi\sum_{j'>j}\kappa_{j}\kappa_{j'}^{*}e^{(j-j')\theta}a_{j'}(t). \label{eq:cascade_hanger_out_a3}
\end{align}
Combining Eqs.\,\eqref{eq:cascade_hanger_in_a3} and \eqref{eq:cascade_hanger_out_a3}, we obtain the input-output relations
\begin{align}
	l_{\rm out} &= l_{\rm in} + \sqrt{2\pi}\kappa_{j'}^{*}\sum_{j'=1}^{N} e^{i(j'-1)\theta}a_{j}, \\
	r_{\rm out} &= r_{\rm in} + \sqrt{2\pi}\kappa_{j'}^{*}\sum_{j'=1}^{N} e^{i(1-j')\theta}a_{j}.
\end{align}

We note that the definition of input and output fields are different from that in Section.\,\ref{sec:cascade_hanger}. To get the exact form, we respectively redefine the operators $l_{\rm in}$ and $r_{\rm in}$ as the input fields at the right and left side of the waveguide, and $l_{\rm out}$ and $r_{\rm out}$ as the output fields at the left and right sides of the waveguide. In other words, we perform the transform $l_{\rm in} \rightarrow e^{i(N-1)\theta}l_{\rm in}$, $r_{\rm out} \rightarrow e^{-i(N-1)\theta}r_{\rm out}$. The new input-output relations are
\begin{align}
	\dot{a}_{j} &= -i\omega_{j}a_{j} 
	-2\pi\sum_{j'=1}^{N}\kappa_{j}\kappa_{j'}^{*}e^{i\left|j'-j\right|\theta} a_{j'}(t)
	-i \sqrt{2\pi}\kappa_{j} l_{\rm in}\left(t\right)e^{-i(j-N)\theta}
	-i \sqrt{2\pi}\kappa_{j} r_{\rm in}\left(t\right)e^{i(j-1)\theta}, \\
	l_{\rm out} &=  e^{i(N-1)\theta}l_{\rm in} + \sum_{j'=1}^{N} \left(\sqrt{\gamma_{j'}}\right)^{*}e^{i(j'-1)\theta}a_{j}, \\
	r_{\rm out} &=  e^{i(N-1)\theta}r_{\rm in} + \sum_{j'=1}^{N} \left(\sqrt{\gamma_{j'}}\right)^{*}e^{i(N-j')\theta}a_{j}.
\end{align}
Correspondingly, the scattering coefficients reads
\begin{align}
	S_{11} &= \frac{\langle l_{\rm out} \rangle}{\langle r_{\rm in}\rangle},\,
	S_{21} =  \frac{\langle r_{\rm out} \rangle}{\langle r_{\rm in} \rangle} 
	\,\text{with}\,\langle l_{\rm in}\rangle = 0,\,\text{and}\,
	S_{12} = \frac{\langle l_{\rm out} \rangle}{\langle l_{\rm in} \rangle},\,
	S_{22} = \frac{\langle r_{\rm out} \rangle}{\langle l_{\rm in}\rangle}
	\,\text{with}\,\langle r_{\rm in}\rangle = 0. 
\end{align}

For the simplicity of calculation, we define
\begin{align}
	c_j &= \sum_{j' \geq j} \left(\sqrt{\gamma_{j'}}\right)^{*}e^{ij'\theta}a_{j'},\,
	d_j = \sum_{j' \leq j} \left(\sqrt{\gamma_{j'}}\right)^{*}e^{-ij'\theta}a_{j'},
\end{align}
such that the input-output relation can be written in a compact form
\begin{align}
	l_{\rm out} &= e^{i(N-1)\theta}l_{\rm in} + e^{-i\theta}c_{1} ,\\
	r_{\rm out} &=  e^{i(N-1)\theta}r_{\rm in} + e^{-i\theta}d_{N}.
\end{align}
On the other hand, the steady-state solution of intra-resonator field, $a_{j}$, reads
\begin{align}
	\left(i\omega_{j} - \gamma_{j} + \frac{\gamma_{a_j}}{2} \right)a_{j} &=
	-e^{-ij\theta}\sqrt{\gamma_j}c_j
	-e^{+ij\theta}\sqrt{\gamma_j}d_j
	-\sqrt{\gamma_j} l_{\rm in}\left(t\right)e^{-i(j-N)\theta}
	-\sqrt{\gamma_j} r_{\rm in}\left(t\right)e^{i(j-1)\theta}, \label{eq:chain_hanger_ass_app}
\end{align}
and thus
\begin{align}
	c_j &= \sum_{j'\geq j}-\frac{\gamma_{j'}}{i\omega_{j'} - \gamma_{j'} + \frac{\gamma_{a_{j'}}}{2}}
	\left(c_{j'} + e^{+i2j'\theta}d_{j'} + e^{+iN\theta}l_{\rm in} + e^{i(2j'-1)\theta}r_{\rm in}\right),\\
	d_j &= \sum_{j'\leq j} -\frac{\gamma_{j'}}{i\omega_{j'} - \gamma_{j'} + \frac{\gamma_{a_{j'}}}{2}}
	\left(e^{-i2j'\theta}c_{j'} + d_{j'} + e^{+i(N-2j')\theta}l_{\rm in} + e^{-i\theta}r_{\rm in}\right).
\end{align}
In these regards, one can readily obtain the input-output relations by solving a linear equation.

\section{Numerical solution for a homogeneous hanger-type-resonator chain}\label{app:cascade_hanger_numerical}
Let us now derive an analytical expression for the scattering coefficients for the special case where $\gamma_j \equiv \gamma$, $\gamma_{a_j} \equiv \gamma_{a}$, $\Delta_{k} \equiv \Delta$. We define
\begin{align}
	c_j &= \sum_{j' \geq j} \left(\sqrt{\gamma_{j'}}\right)^{*}e^{ij'\theta}a_{j'},\,
	d_j = \sum_{j' \leq j} \left(\sqrt{\gamma_{j'}}\right)^{*}e^{-ij'\theta}a_{j'}, \label{eq:chain_hanger_cd_app}
\end{align}
such that the steady-state solution of Eq.\,\eqref{eq:chain_hanger_ass_app} can be written as
\begin{align}
	a_{j} &=
	-\frac{e^{-ij\theta}\sqrt{\gamma_j}}{i\omega_{j} - \gamma_{j} + \frac{\gamma_{a_j}}{2}}c_j
	-\frac{e^{+ij\theta}\sqrt{\gamma_j}}{i\omega_{j} - \gamma_{j} + \frac{\gamma_{a_j}}{2}}d_j
	-\frac{e^{-i(j-N)\theta}\sqrt{\gamma_j}}{i\omega_{j} - \gamma_{j} + \frac{\gamma_{a_j}}{2}}l_{\rm in}
	-\frac{e^{i(j-1)\theta}\sqrt{\gamma_j}}{i\omega_{j} - \gamma_{j} + \frac{\gamma_{a_j}}{2}}r_{\rm in}. \label{eq:chain_hanger_ass_cd_app}
\end{align}

For convenience, we define $x=\gamma/\left(i\omega - \gamma + \gamma_{a}/2\right)$. By combing Eqs.\,\eqref{eq:chain_hanger_cd_app} and \eqref{eq:chain_hanger_ass_cd_app}, we obtain a set of linear equations
\begin{align}
	xe^{-i2\theta}c_{1} + \left(x+1\right)d_{1} 
	+ x\left(e^{+i(N-2)\theta}l_{\rm in}
	+e^{-i\theta}r_{\rm in}\right) &= 0, \\
	xe^{-i2(j+1)\theta}c_{j+1} 
	-d_{j} + \left(x+1\right)d_{j+1} 
	+ x\left(e^{+i(N-2(j+1))\theta}l_{\rm in}
	+e^{-i\theta}r_{\rm in}\right) &= 0, \\
	\left(x+1\right)c_{j} - c_{j+1} 
	+ xe^{+i2j\theta}d_{j}
	+ x\left(e^{+iN\theta}l_{\rm in}
	+e^{i(2j-1)\theta}r_{\rm in}\right) &= 0,\\
	\left(x+1\right)c_{N} + xe^{+i2N\theta}d_{N}
	+ x\left(e^{+iN\theta}l_{\rm in}
	+e^{i(2N-1)\theta}r_{\rm in}\right) &= 0.
\end{align}
Next, we decouple the variables $c_j$ and $d_j$ and obtain the following two sets of linear equations
\begin{align}
	-\frac{\left(1+2x\right)}{x}c_{1}
	+\frac{\left(1+x\right)}{x}c_{2} 
	- e^{iN\theta}l_{\rm in}
	- e^{i\theta}r_{\rm in} &= 0, \\
	\frac{\left(x+1\right)e^{i2\theta}}{x}c_{j-1} 
	- \frac{\left(1+2x+e^{i2\theta}\right)}{x}c_{j} 
	+\frac{\left(1+x\right)}{x}c_{j+1} 
	+ e^{iN\theta}\left(e^{i2\theta}-1\right)l_{\rm in} &= 0,\, j=2,\cdots,N-1, \\
	\frac{\left(x+1\right)e^{i2\theta}}{x}c_{N-1} 
	- \frac{\left(1+2x+e^{i2\theta}\right)}{x}c_{N} 
	+ e^{iN\theta}\left(e^{i2\theta}-1\right)l_{\rm in} &= 0.
\end{align}
\begin{align}
	-\frac{\left(1+e^{i2\theta}+2x\right)}{x}d_{1} 
	+ \frac{e^{i2\theta}\left(x+1\right)}{x}d_{2}
	+ \left(e^{i\theta}
	- e^{-i\theta}\right)r_{\rm in} &= 0,\\
	\frac{\left(1+x\right)}{x}d_{j-1}
	-\frac{\left(1+e^{i2\theta}+2x\right)}{x}d_{j} 
	+ \frac{e^{i2\theta}\left(x+1\right)}{x}d_{j+1}
	+ \left(e^{i\theta}
	- e^{-i\theta}\right)r_{\rm in} &= 0,\,j=2,\cdots,N-1,\\
	\frac{\left(1+x\right)}{x}d_{N-1}
	-\frac{\left(1+2x\right)}{x}d_{N} 
	- e^{-iN\theta}l_{\rm in}
	- e^{-i\theta}r_{\rm in} &= 0.
\end{align}

The above equations consist of tridiagonal matrices, which can be efficiently solved by using the so-called Thomas or TDMA algorithm. That is
$c_N = b'_N$, $c_j = b'_j - a'_j c_{j+1}$ for $j=N-1,\cdots,1$, where
\begin{align}
	a'_j = \begin{cases}
	-\frac{1+x}{1+2x},\,j=1,\\
	-\frac{1+x}{\left(1+2x+e^{i2\theta}\right) 
	+ a'_{j-1}(1+x)e^{i2\theta}},\,j=2,\cdots,N-1,
\end{cases},\,
	b'_j = \begin{cases}
	-\frac{x\left(e^{iN\theta}l_{\rm in}
	+ e^{i\theta}r_{\rm in}\right)}{1+2x},\,j=1,\\
	-\frac{ x e^{iN\theta}\left(1-e^{i2\theta}\right)l_{\rm in}-b'_{j-1}(1+x)e^{i2\theta}}{\left(1+2x+e^{i2\theta}\right) 
	+ a'_{j-1}(1+x)e^{i2\theta}},\,j=2,\cdots,N-1.
\end{cases}
\end{align}
Or, $d_1 = b'_N$, $d_{N-j+1} = b'_{j} - a'_{j} d_{N-j}$ for $j=N-1,\cdots,1$, where
\begin{align}
	a'_j = \begin{cases}
	-\frac{1+x}{1+2x},\,j=1,\\
	-\frac{1+x}{\left(1+2x+e^{i2\theta}\right) 
	+ a'_{j-1}(1+x)e^{i2\theta}},\,j=2,\cdots,N-1,
\end{cases},\,
	b'_j = \begin{cases}
	-\frac{x\left(e^{-iN\theta}l_{\rm in}
	+ e^{-i\theta}r_{\rm in}\right)}{1+2x},\,j=1,\\
	-\frac{ x \left(e^{-i\theta}-e^{i\theta}\right)r_{\rm in}-b'_{j-1}(1+x)e^{i2\theta}}{\left(1+2x+e^{i2\theta}\right) 
	+ a'_{j-1}(1+x)e^{i2\theta}},\,j=2,\cdots,N-1
\end{cases}
\end{align}

\section{A chain of necklace-type resonators with two boundary conditions}\label{app:cascade_necklace}
For hard-wall boundary conditions, we define
\begin{align}
	c_1 &= \sum_{k'=1}^{N}\sqrt{\gamma_{1,k'}}a_{k'},\,
	c_2 = \sum_{k'=1}^{N}(-1)^{k'}\sqrt{\gamma_{2,k'}}a_{k'}, \label{eq:chain_hanger_c_app}
\end{align}
such that
\begin{align}
	a_{k} = - \frac{\frac{1}{2}\sqrt{\gamma_{1,k}}c_1 + \frac{(-1)^{k}}{2}\sqrt{\gamma_{2,k}}c_2 
	+ \sqrt{\gamma_{1,k}}b_{1,{\rm in}} + (-1)^{k}\sqrt{\gamma_{2,k}} b_{2,{\rm in}}}{i\Delta_{k}+\frac{\gamma_{a}}{2}}. \label{eq:chain_hanger_ass_c_app}
\end{align}
Combing Eqs.\,\eqref{eq:chain_hanger_c_app} and \eqref{eq:chain_hanger_ass_c_app}, we obtain
\begin{align}
	c_1 &= -\sum_{k=1}^{N}\frac{\frac{1}{2}\gamma_{1,k}c_1 + \frac{(-1)^{k}}{2}\sqrt{\gamma_{1,k}\gamma_{2,k}}c_2 
	+ \gamma_{1,k}b_{1,{\rm in}} + (-1)^{k}\sqrt{\gamma_{1,k}\gamma_{2,k}} b_{2,{\rm in}}}{i\Delta_{k}+\frac{\gamma_{a}}{2}},\\
	c_2 &= -\sum_{k=1}^{N}\frac{\frac{(-1)^{k}}{2}\sqrt{\gamma_{1,k}\gamma_{2,k}}c_1 
	+ \frac{1}{2}\gamma_{2,k} c_2 
	+ (-1)^{k}\sqrt{\gamma_{1,k}\gamma_{2,k}}b_{1,{\rm in}} 
	+ \gamma_{2,k} b_{2,{\rm in}}}{i\Delta_{k}+\frac{\gamma_{a}}{2}}.
\end{align}
The solutions of $c_1$ and $c_2$ reads
\begin{align}
	c_1 &= -\frac{\left[\left(1+\frac{1}{2}\sum_{k=1}^{N}\frac{\gamma_{2,k}}{i\Delta_k+\frac{\gamma_a}{2}}\right)
	\left(\sum_{k=1}^{N}\frac{\gamma_{1,k}}{i\Delta_k+\frac{\gamma_a}{2}}\right)
	-\frac{1}{2}\left(\sum_{k=1}^{N}\frac{(-1)^{k}\sqrt{\gamma_{1,k}\gamma_{2,k}}}{i\Delta_{k}+\frac{\gamma_{a}}{2}}\right)^2\right]
	b_{1,{\rm in}}
	+ \left(\sum_{k=1}^{N}\frac{(-1)^{k}\sqrt{\gamma_{1,k}\gamma_{2,k}}}{i\Delta_k+\frac{\gamma_a}{2}}\right)
	b_{2,{\rm in}}}
	{\left(1+\frac{1}{2}\sum_{k=1}^{N}\frac{\gamma_{1,k}}{i\Delta_{k}+\frac{\gamma_{a}}{2}}\right)
	\left(1+\frac{1}{2}\sum_{k=1}^{N}\frac{\gamma_{2,k}}{i\Delta_{k}+\frac{\gamma_{a}}{2}}\right)
	-\left(\frac{1}{2}\sum_{k=1}^{N}\frac{(-1)^{k}\sqrt{\gamma_{1,k}\gamma_{2,k}}}{i\Delta_{k}+\frac{\gamma_{a}}{2}}\right)^2},\\
	c_2 &= -\frac{\left(\sum_{k=1}^{N}\frac{(-1)^{k}\sqrt{\gamma_{1,k}\gamma_{2,k}}}{i\Delta_k+\frac{\gamma_a}{2}}\right)
	b_{1,{\rm in}}
	+\left[\left(1+\frac{1}{2}\sum_{k=1}^{N}\frac{\gamma_{1,k}}{i\Delta_k+\frac{\gamma_a}{2}}\right)
	\left(\sum_{k=1}^{N}\frac{\gamma_{2,k}}{i\Delta_k+\frac{\gamma_a}{2}}\right)
	-\frac{1}{2}\left(\sum_{k=1}^{N}\frac{(-1)^{k}\sqrt{\gamma_{1,k}\gamma_{2,k}}}{i\Delta_{k}+\frac{\gamma_{a}}{2}}\right)^2\right]
	b_{2,{\rm in}}}
	{\left(1+\frac{1}{2}\sum_{k=1}^{N}\frac{\gamma_{1,k}}{i\Delta_{k}+\frac{\gamma_{a}}{2}}\right)
	\left(1+\frac{1}{2}\sum_{k=1}^{N}\frac{\gamma_{2,k}}{i\Delta_{k}+\frac{\gamma_{a}}{2}}\right)
	-\left(\frac{1}{2}\sum_{k=1}^{N}\frac{(-1)^{k}\sqrt{\gamma_{1,k}\gamma_{2,k}}}{i\Delta_{k}+\frac{\gamma_{a}}{2}}\right)^2}.
\end{align}
Thus, the input-output relations can be readily obtained by inserting $c_1$ and $c_2$ into Eqs.\,\eqref{eq:inout_a_app}-\eqref{eq:inout_b_app}.

For periodic boundary conditions, we define
\begin{align}
	c_1 &= \sum_{k'=1}^{N}\sqrt{\gamma_{1,k'}}a_{k'},\,
	c_2 = -\sum_{k'=1}^{N}\sqrt{\gamma_{2,k'}}a_{k'}, \label{eq:chain_hanger_c_app2}
\end{align}
such that
\begin{align}
	a_{k} = - \frac{\frac{1}{2}\sqrt{\gamma_{1,k}}c_1 - \frac{1}{2}\sqrt{\gamma_{2,k}}c_2 
	+ \sqrt{\gamma_{1,k}}b_{1,{\rm in}} - \sqrt{\gamma_{2,k}} b_{2,{\rm in}}}{i\Delta_{k}+\frac{\gamma_{a}}{2}}. \label{eq:chain_hanger_ass_c_app2}
\end{align}
Combing Eqs.\,\eqref{eq:chain_hanger_c_app2} and \eqref{eq:chain_hanger_ass_c_app2}, we obtain
\begin{align}
	c_1 &= -\sum_{k=1}^{N}\frac{\frac{1}{2}\gamma_{1,k}c_1 - \frac{1}{2}\sqrt{\gamma_{1,k}\gamma_{2,k}}c_2 
	+ \gamma_{1,k}b_{1,{\rm in}} - \sqrt{\gamma_{1,k}\gamma_{2,k}} b_{2,{\rm in}}}{i\Delta_{k}+\frac{\gamma_{a}}{2}},\\
	c_2 &= -\sum_{k=1}^{N}\frac{-\frac{1}{2}\sqrt{\gamma_{1,k}\gamma_{2,k}}c_1 
	+ \frac{1}{2}\gamma_{2,k} c_2 
	-\sqrt{\gamma_{1,k}\gamma_{2,k}}b_{1,{\rm in}} 
	+ \gamma_{2,k} b_{2,{\rm in}}}{i\Delta_{k}+\frac{\gamma_{a}}{2}}.
\end{align}
The solutions of $c_1$ and $c_2$ reads
\begin{align}
	c_1 &= \frac{-\left[\left(1+\frac{1}{2}\sum_{k=1}^{N}\frac{\gamma_{2,k}}{i\Delta_k+\frac{\gamma_a}{2}}\right)
	\left(\sum_{k=1}^{N}\frac{\gamma_{1,k}}{i\Delta_k+\frac{\gamma_a}{2}}\right)
	-\frac{1}{2}\left(\sum_{k=1}^{N}\frac{\sqrt{\gamma_{1,k}\gamma_{2,k}}}{i\Delta_{k}+\frac{\gamma_{a}}{2}}\right)^2\right]
	b_{1,{\rm in}}
	+ \left(\sum_{k=1}^{N}\frac{\sqrt{\gamma_{1,k}\gamma_{2,k}}}{i\Delta_k+\frac{\gamma_a}{2}}\right)
	b_{2,{\rm in}}}
	{\left(1+\frac{1}{2}\sum_{k=1}^{N}\frac{\gamma_{1,k}}{i\Delta_{k}+\frac{\gamma_{a}}{2}}\right)
	\left(1+\frac{1}{2}\sum_{k=1}^{N}\frac{\gamma_{2,k}}{i\Delta_{k}+\frac{\gamma_{a}}{2}}\right)
	-\left(\frac{1}{2}\sum_{k=1}^{N}\frac{\sqrt{\gamma_{1,k}\gamma_{2,k}}}{i\Delta_{k}+\frac{\gamma_{a}}{2}}\right)^2},\\
	c_2 &= \frac{\left(\sum_{k=1}^{N}\frac{\sqrt{\gamma_{1,k}\gamma_{2,k}}}{i\Delta_k+\frac{\gamma_a}{2}}\right)
	b_{1,{\rm in}}
	-\left[\left(1+\frac{1}{2}\sum_{k=1}^{N}\frac{\gamma_{1,k}}{i\Delta_k+\frac{\gamma_a}{2}}\right)
	\left(\sum_{k=1}^{N}\frac{\gamma_{2,k}}{i\Delta_k+\frac{\gamma_a}{2}}\right)
	-\frac{1}{2}\left(\sum_{k=1}^{N}\frac{\sqrt{\gamma_{1,k}\gamma_{2,k}}}{i\Delta_{k}+\frac{\gamma_{a}}{2}}\right)^2\right]
	b_{2,{\rm in}}}
	{\left(1+\frac{1}{2}\sum_{k=1}^{N}\frac{\gamma_{1,k}}{i\Delta_{k}+\frac{\gamma_{a}}{2}}\right)
	\left(1+\frac{1}{2}\sum_{k=1}^{N}\frac{\gamma_{2,k}}{i\Delta_{k}+\frac{\gamma_{a}}{2}}\right)
	-\left(\frac{1}{2}\sum_{k=1}^{N}\frac{\sqrt{\gamma_{1,k}\gamma_{2,k}}}{i\Delta_{k}+\frac{\gamma_{a}}{2}}\right)^2}.
\end{align}
Similarily, the input-output relations can be readily obtained by inserting $c_1$ and $c_2$ into Eqs.\,\eqref{eq:inout_a_app}-\eqref{eq:inout_b_app}.

\end{widetext}
\bibliography{PT_ref}  
\end{document}